\documentclass[aps,prd,twocolumn,showpacs]{revtex4}

\usepackage[usenames,dvipsnames]{color}

\usepackage{graphicx}
\usepackage{amssymb}

\newcommand{\C}[6]{{C_{#1}^{#2}{}_{#3}^{#4}{}_{#5}^{#6}}}
\newcommand{\E}[7]{{E_{#1}^{#4}{}_{#2}^{#5}{}_{#3}^{#6}
{}^{}_{#7}}}

\newcommand{\h}[1]{{}^{\mbox{\,\tiny $\{#1\}\!$}}h}
\newcommand{\pert}[2]{{}^{\mbox{\,\tiny $\{#1\}\!$}}{#2}}
\newcommand{\T}[1]{{}^{\mbox{\,\tiny $\{#1\}\!$}}T}
\newcommand{\Ph}[1]{{}^{\mbox{\,\tiny $\{#1\}\!$}}\Phi}
\newcommand{\sign}[2]{{{}^{\mbox{\tiny $({#1})\!$}}{#2}}}
\newcommand{\Source}[7]{{{}^{\mbox{\tiny $({#1})\!$}}S_{#2}^{#3}
{}_{#4}^{#5}{}_{#6}^{#7}}}
\newcommand{\tildeSource}[7]{{{}^{\mbox{\tiny $({#1})\!$}}
{\tilde S}_{#2}^{#3}{}_{#4}^{#5}{}_{#6}^{#7}}}
\newcommand{\hblue}{{\bar h}}
\newcommand{\hred}{{\hat h}}
\newcommand{\Hblue}{{\bar H}}
\newcommand{\Hred}{{\hat H}}
\newcommand{\Kblue}{{\bar K}}
\newcommand{\Kred}{{\hat K}}
\newcommand{\lblue}{{\bar l}}
\newcommand{\lred}{{\hat l}}
\newcommand{\mblue}{{\bar m}}
\newcommand{\mred}{{\hat m}}
\newcommand{\Pblue}{{\bar\Pi}}
\newcommand{\Pred}{{\hat\Pi}}
\newcommand{\source}[7]{{{}^{\mbox{\tiny $({#1})\!$}}
{\cal I}_{#2}^{#3}{}_{#4}^{#5}{}_{#6}^{#7}}}
\newcommand{\psired}{{\hat\psi}}
\newcommand{\psiblue}{{\bar\psi}}
\newcommand{\Psired}{{\hat\Psi}}
\newcommand{\Psiblue}{{\bar\Psi}}
\newcommand{\Psitildered}{{\hat{\tilde\Psi}}}
\newcommand{\Psitildeblue}{{\bar{\tilde\Psi}}}

\begin{document}

\title{Second and higher-order perturbations of a spherical
spacetime}

\author{David Brizuela}
\author{Jos\'e M. Mart\'{\i}n-Garc\'{\i}a}
\author{Guillermo A. Mena Marug\'an}
\affiliation{Instituto de Estructura de la Materia, CSIC, Serrano
121-123, 28006 Madrid, Spain}

\date{\today}


\begin{abstract}
The Gerlach and Sengupta (GS) formalism of coordinate-invariant,
first-order, spherical and nonspherical perturbations around an
arbitrary spherical spacetime is generalized to higher orders,
focusing on second-order perturbation theory. The GS harmonics are
generalized to an arbitrary number of indices on the unit sphere and
a formula is given for their products. The formalism is optimized
for its implementation in a computer algebra system, something that
becomes essential in practice given the size and complexity of the
equations. All evolution equations for the second-order
perturbations, as well as the conservation equations for the
energy-momentum tensor at this perturbation order, are given in
covariant form, in Regge-Wheeler gauge.
\end{abstract}



\maketitle


\section{Introduction}
\label{intro}

When a physical problem cannot be solved exactly, one usually
appeals to approximate methods. Perturbation theory provides one
such approximate approach, allowing a description in terms of small
departures around an exact solution. In the context of General
Relativity, perturbation theory plays a prominent role in analyzing
and understanding dynamical processes, being nowadays an efficient
and natural complement to full numerical relativity simulations
\cite{Seidel}.

In particular, perturbation theory is used to study the stability
properties of solutions of interest: black hole spacetimes
\cite{Chandra}, cosmological solutions \cite{Cosmo}, critical
solutions \cite{carsten}, and many other. It also allows us to check
the presence of gauge instabilities \cite{KN}, constraint violations
\cite{constraintdamping}, and other types of instabilities in
the various formulations of the Einstein equations implemented in
numerical relativity, since numerical errors can be considered
themselves as distortions of the solution that one is computing. Of
outmost importance, perturbation theory can provide us with
estimates of the amount of gravitational radiation and of the signal
profiles emitted in astrophysical scenarios like an oscillating
neutron star \cite{Nils}, the gravitational collapse of a star
\cite{Harada}, an extreme mass-ratio binary \cite{Drasco}, or a
close limit head-on collision of two black holes \cite{GNP96}.

The complexity of the expressions involved in the perturbation of
the equations of General Relativity is very high, rapidly increasing
when working at higher orders of perturbation. Most of the previous
investigations have been carried out at first order, and using
highly symmetric unperturbed solutions, what simplifies the problem.
Simple backgrounds have been perturbed at second-order level: for
example, the Friedmann-Robertson-Walker spacetime \cite{BMT02}, the
Kerr spacetime \cite{CaLo99}, or the Schwarzschild spacetime
\cite{PrGa00, GNP96*}. The change in the oscillation modes of a
stationary star owing to its rotation has been studied in Refs.
\cite{CF91,KOJ92}, regarding the rotation as a first-order
perturbation of a static star (and therefore the change as a
second-order effect). The critical exponent of angular momentum
scaling has been predicted for scalar field collapse using
second-order perturbation arguments \cite{ang99}. Special credit
deserves the seminal work of Cunningham, Price, and Moncrief in 1980
where the second-order nonspherical perturbations of a collapsing
star of dust were studied including the problem of matching of the
internal and external perturbations through the surface of the star
\cite{CPM80}. Second-order perturbations of the general problem of
matching through a surface have been analyzed in Ref. \cite{Mars05}.

Different reasons justify the importance of going beyond first-order
perturbation theory. First, one must confront the obvious desire to
reach higher accuracy in the numerical simulations of
perturbative approximations to self-gravitating systems. In
addition, second-order perturbations can estimate quantitatively the
range of applicability of the first-order results by providing error
estimates. Besides, they should enable us to study and interpret the
nonlinearity of General Relativity in terms of the coupling among
first-order modes \cite{PBGNS06}, a study that might allow us to
model secular interactions which are too slow to be followed by
using full numerical simulations or which might be incorrectly
interpreted as small numerical errors.

In this work, we construct a generic framework to analyze second and
higher-order spherical and nonspherical perturbations of an
arbitrary spherical spacetime, without restricting to any particular
matter model, and show how to deal with the general case while
keeping a still manageable size for the resulting expressions. This
work can be considered as a continuation of the work of Gerlach and
Sengupta (GS) at first order \cite{GeSe79, GeSe80}, a formalism
which was revived in Ref. \cite{critfluid} and is currently
considered optimal for perturbations of generic spherical
spacetimes \cite{NaRe05,MaPo05} (even though for specific
matter models the formalism might admit further simplifications).
For example, its application to a perfect fluid spacetime with a
two-parameters equation of state was considered in Ref.
\cite{fluidpert1}, including the matching of the fluid perturbations
to an exterior spacetime through a moving timelike surface
\cite{fluidpert2}.

The GS formalism is based on four basic ingredients: a 2+2
decomposition of the spacetime separating the spherical $S^2$
symmetry orbits from a general 1+1 Lorentzian manifold $M^2$; the
use of a covariant description both on the Lorentzian manifold and
on the 2-sphere; the decomposition of the perturbations in $S^2$
tensor harmonics; and the use of gauge-invariant perturbation
variables. The formalism developed in the present paper makes use of
the first three ingredients, leaving the construction of
gauge-invariants for a future work. The use of a covariant
notation is particularly convenient: on the one hand, it allows us
to formulate all equations without choosing coordinates on $M^2$,
something that becomes very useful on dynamical backgrounds; on the
other hand, it eliminates all trigonometric factors from the
equations of motion, factors which do not contain any relevant
information and typically obscure the geometrical interpretation of
the results. The use of covariant notation, however, makes
computations more complicated. To overcome this problem we will
intensively use computer algebra tools, specially designed and
developed by us for abstract tensor computations. All equations in
this paper have been programmed and checked with those tools,
resulting in a computer framework which allows us to work
efficiently with applications of the formalism of high-order
perturbations.

For some matter contents it is possible to further simplify the
perturbative formalism by constructing scalar combinations of the
perturbations which encode the purely dynamical degrees of freedom
and obey evolution equations free of constraints (the so-called {\em
master equations} for the {\em master scalars}). Among other matter
models, this is possible for vacuum \cite{ReWh57,Zeri70} and for the
Maxwell field \cite{MoncriefRN}, but such master equations have not
been found, for instance, for the case of a scalar field. Those
scalars can be given in a gauge-invariant form \cite{Moncrief74} and
expressed in covariant form within the GS framework
\cite{GeSe79,SaTi01}. In this article we will concentrate on the
construction of a generic perturbative formalism that can be applied
to any background spherical spacetime. The construction of
gauge-invariant master scalars at high perturbative orders will be
the subject of a future work.

The rest of this article is organized as follows. Section II
introduces the fundamental concepts and equations of perturbation
theory, and provides closed formulas for the $n$th-order
perturbations of the geometric quantities of interest (formulas that
are new in the literature to the best of our knowledge). The
notation used for the spherical background manifold is explained in
Section III. Section IV introduces the Regge-Wheeler-Zerilli
harmonics \cite{ReWh57,Zeri70b} and generalize them to arbitrary rank
tensors. Formulas for their products are given, based on the
representation matrices of the rotation group. Basically, nothing in
that section is new, but it has been conveniently recast into GS
notation, which proves to be very useful for applications and
computational efficiency. Nonspherical perturbations are discussed
in Section V, giving for the first time complete sources for the
evolution equations of second-order perturbations of a general
spherical background, as well as sources for the energy-momentum
conservation equations. Section VI contains a summary of our results
and further discussions. Finally, several appendices are added. They
explain different aspects of the definition of spherical functions,
the symmetric trace-free part of tensors, and the pure-orbital
harmonics. They also provide the GS equations for
first-order perturbation theory and describe the procedure followed
to implement our calculations in {\it Mathematica} \cite{Mathematica}.


\section{Perturbation theory in General Relativity}
\label{GRpert}

In General Relativity, one has to solve ten coupled nonlinear
partial differential equations for the metric. Using perturbation
theory, the problem can be reformulated as an infinite hierarchy of
linear differential equations for the modification of the metric
with respect to a known solution. Furthermore, in this hierarchy of
equations, the principal parts are always given by the same
differential operator acting on a perturbative correction of
increasing order. In this way, one translates the difficulty from
nonlinearity to the infinite number of equations. The key assumption
of the perturbative scheme is that one can truncate the problem at a
finite order and still obtain an approximate solution to the
original system.

In order to introduce this perturbative hierarchy, let us start by
considering an $\epsilon$-family of 4-dimensional metrics
$\tilde{g}_{\mu\nu}(\epsilon)$ on a certain manifold ${\cal M}$,
where $\epsilon$ is a dimensionless parameter \cite{gen}. The matter
field content of these spacetimes provides another $\epsilon$-family
which will be denoted abstractly by $\tilde\Phi(\epsilon)$. For
simplicity, we assume that these families are smooth in the
parameter $\epsilon$ (or at least $C^n$ with a locally Lipschitz
$n$-th derivative if we are interested only in perturbation theory
up to order $n>0$). In particular, the metric and matter fields can
be expanded as \cite{JLM}
\begin{eqnarray} \label{metricexpansion}
\tilde{g}_{\mu\nu}(\epsilon) &=& g_{\mu\nu} +
\sum_{n=1}^\infty \frac{\epsilon^n}{n!} \h{n}_{\mu\nu} , \\
\tilde{\Phi}(\epsilon) &=& \Phi + \sum_{n=1}^\infty
\frac{\epsilon^n}{n!} \Ph{n}.
\end{eqnarray}
The $\epsilon=0$ fields $g_{\mu\nu}$ and $\Phi$ will be referred to
as the ``background'' metric and matter. To distinguish the
background objects from their ``perturbed'' $\epsilon\neq 0$
counterparts, we will denote the latter with a tilde
$(\,\tilde{}\,)$.

The coefficients $\h{n}_{\mu\nu}$ and $\Ph{n}$ are tensors on the
manifold ${\cal M}$ \cite{regu}, and in what follows their indices
are lowered or raised with the background metric $g_{\mu\nu}$ and
its inverse $g^{\mu\nu}$. They can be obtained by repeated
differentiation of the perturbed objects with respect to $\epsilon$.
Actually, it proves most convenient to introduce a formal
``perturbation'' operator $\Delta$ with the properties of a
derivative, so that any object $\tilde{T}(\epsilon)$ can be expanded
as
\begin{equation}
\tilde{T}(\epsilon) = T + \sum_{n=1}^\infty \frac{\epsilon^n}{n!}
\Delta^n[T].
\end{equation}
For instance, $\Delta[g_{\mu\nu}]=\h{1}_{\mu\nu}$ and
$\Delta[\h{n}_{\mu\nu}]= \h{n+1}_{\mu\nu}$. In this notation, the
brackets are intended to avoid confusion with index positioning
because, e.g., for a given vector $v^\mu$:
\begin{equation}
\Delta[v_\mu] = \Delta[g_{\mu\nu}v^\nu] =
g_{\mu\nu}\Delta[v^\nu] + \h{1}_{\mu\nu}v^\nu
\not= g_{\mu\nu}\Delta[v^\nu],
\end{equation}
so that the notation $\Delta v_\mu$ might be misleading.

The expansion for the inverse metric can be obtained by iteration of
the identity
\begin{equation}
\tilde{g}^{\mu\nu}\equiv g^{\mu\nu}- g^{\mu\lambda}
(\tilde{g}_{\lambda\sigma}-g_{\lambda\sigma})\tilde{g}^{\sigma\nu}
\end{equation}
or, equivalently, by repeated perturbation of the relation
\begin{equation}
\Delta[g^{\mu\nu}]=
-g^{\mu\alpha}\Delta[g_{\alpha\beta}]g^{\beta\nu}.
\end{equation}
This leads to
\begin{equation}
\tilde{g}^{\mu\nu}=g^{\mu\nu} - \epsilon \h{1}{}^{\mu\nu} -
\frac{\epsilon^2}{2} (\h{2}{}^{\mu\nu} -
 2\h{1}{}^{\mu\alpha}\h{1}_\alpha{}^\nu) +
O(\epsilon^3),
\end{equation}
and, more generally, to a perturbation of the form
\begin{eqnarray} \label{generalgn}
\Delta^n[g^{\mu\nu}] &=&
\sum_{(k_i)} (-1)^m\frac{n!}{k_1!\,...\,k_m!} \\
&\times& \h{k_m}^{\mu\alpha}\;\h{k_{m-1}}_{\alpha\beta}\;...\;
\h{k_2}_{\tau\rho}\;\h{k_1}^{\rho\nu} , \nonumber
\end{eqnarray}
where the sum extends to the $2^{n-1}$ sorted partitions of $n$ in
$m\le n$ positive integers $k_1+...+k_m=n$. For example, for $n=4$
there are eight partitions: $(4)$, $(1,3)$, $(3,1)$, $(2,2)$,
$(1,1,2)$, $(1,2,1)$, $(2,1,1)$, and $(1,1,1,1)$.

The Christoffel symbols are
\begin{eqnarray} \label{Gamma1}
\tilde\Gamma^\alpha{}_{\mu\nu} &=& \Gamma^\alpha_{\mu\nu} +
\epsilon \; \h{1}^\alpha{}_{\mu\nu} \\
&+& \frac{\epsilon^2}{2}
\left(\h{2}^\alpha{}_{\mu\nu}
- 2 \h{1}^{\alpha\beta}\h{1}_{\beta\mu\nu}\right)
+ O(\epsilon^3),
\nonumber
\end{eqnarray}
where we have defined the three-indices perturbation
\begin{equation} \label{Gamma2}
\h{n}_{\alpha\mu\nu} \equiv \frac{1}{2}\left( \h{n}_{\alpha\mu ;\nu}
+ \h{n}_{\alpha\nu ;\mu} - \h{n}_{\mu\nu ;\alpha}\right),
\end{equation}
which is symmetric in its last two indices and satisfies
$\h{0}_{\alpha\mu\nu}=0$. The covariant derivative in (\ref{Gamma2})
is that associated with the background metric. Higher-order terms of
the expansion can be easily computed noting that
\begin{equation} \label{Gamma3}
\Delta\left[\h{n}_{\alpha\mu\nu}\right] =\h{n+1}_{\alpha\mu\nu} -
\h{n}_\alpha{}^\beta \h{1}_{\beta\mu\nu} ,
\end{equation}
which, for $n>1$, leads to
\begin{eqnarray} \label{Gamma4}
\Delta^n[\Gamma^\alpha{}_{\mu\nu}] &=&
\sum_{(k_i)} (-1)^{m+1}\frac{n!}{k_1!\,...\,k_m!} \\
&\times&  \h{k_m}^{\alpha\beta}\;\h{k_{m-1}}_{\beta\gamma}\;...\;
\h{k_{2}}^{\tau\rho}\;\h{k_1}_{\rho\mu\nu}. \nonumber
\end{eqnarray}
Here, the sum extends again to all sorted partitions of $n$. The
case $n=1$ is special, with
$\Delta[\Gamma^\alpha{}_{\mu\nu}]=\h{1}^{\alpha}{}_{\mu\nu}$. Note
that each term in the above expression contains one and only one
tensor $\h{k}_{\rho\mu\nu}$, but a variable number of metric
perturbations.

The perturbations of the Riemann tensor are given by
\begin{eqnarray}
\tilde{R}_{\mu\nu\sigma}{}^\lambda &=& {R}_{\mu\nu\sigma}{}^\lambda
+ 2 \epsilon \;\h{1}^\lambda{}_{\sigma[\mu;\nu]} \\
&+& \epsilon^2
\left(
    \h{2}^\lambda{}_{\sigma[\mu;\nu]}
- 2 \h{1}^{\lambda\alpha}
    \h{1}_{\alpha\sigma[\mu;\nu]} \right. \nonumber \\
&& \qquad \left.
+ 2 \h{1}_{\alpha[\mu}{}^\lambda
    \h{1}^\alpha{}_{\nu]\sigma}
\right)
+ O(\epsilon^3) ,
\nonumber
\end{eqnarray}
with the general term
\begin{eqnarray} \label{pertRiemann}
&& \Delta^n[R_{\mu\nu\alpha}{}^\beta] = \\
&&\quad\nabla_\nu\left(\Delta^n[\Gamma^\beta{}_{\alpha\mu}]\right)
- \sum_{k=1}^{n-1} \left(\matrix{ n \cr k}\right)
\Delta^k[\Gamma^\beta{}_{\lambda\mu}]
\Delta^{n-k}[\Gamma^\lambda{}_{\nu\alpha}] \nonumber \\
&& \quad - \quad (\mu\leftrightarrow\nu) . \nonumber
\end{eqnarray}
This expression does not involve the metric directly. That is, it
only contains the background connection $\nabla$, and the
perturbations of its Christoffel symbols, without assuming that
either the background or the perturbed connections derive from a
metric. Hence, it can be applied to the Palatini equations, for
example. If the background connection derives from a metric, Eqs.
(\ref{Gamma1}--\ref{Gamma4}) ensure that the perturbed connection
also derives from a metric. Then, the (implicit) triple sum in Eq.
(\ref{pertRiemann}) can be rearranged as follows:
\begin{eqnarray} \label{pertRiemann2}
&&\Delta^n[R_{\mu\nu\alpha}{}^\beta] =
\sum_{(k_i)} (-1)^m \frac{n!}{k_1!...k_m!} \\
&&\qquad\times \left[
\h{k_m}^{\beta\lambda_m}...\h{k_2}^{\lambda_3\lambda_2}
\h{k_1}_{\lambda_2\alpha\nu;\mu} \right. \nonumber \\
&&\qquad + \sum_{s=2}^{m}\nonumber \h{k_m}^{\beta\lambda_m}...
\h{k_{s+1}}^{\lambda_{s+2}\lambda_{s+1}}
\h{k_s}_{\lambda_s\lambda_{s+1}\mu}\nonumber \\
&&\qquad\times\left. \h{k_{s-1}}^{\lambda_s\lambda_{s-1}}...
\h{k_2}^{\lambda_3\lambda_2}
\h{k_1}_{\lambda_2\nu\alpha}\right] \nonumber \\
&&\qquad - (\mu\leftrightarrow\nu). \nonumber
\end{eqnarray}

This formula is surprisingly simple because all covariant
derivatives of the metric perturbations are grouped in $h$-terms of
the form (\ref{Gamma2}). It is clear that the explicit sum in
Eq. (\ref{pertRiemann}) contains only that kind of terms; however
the derivatives of the perturbations of the Christoffel symbols give
rise to isolated covariant derivatives of the metric perturbations.
Nonetheless, they can be combined with the sum to obtain the
displayed result. Note also that this expression is already
optimally simplified: all terms in the sums are generically
different.

From the above formulas, we can compute the general perturbation of
the Ricci and Einstein tensors. For Ricci, the perturbation is
obtained from Eq. (\ref{pertRiemann2}) by contracting the $\beta$
and $\nu$ indices. The resulting expression is symmetric in $\alpha$
and $\mu$ owing to the identity
\begin{equation}
\sum_{k=1}^{n}\T{k}^{\alpha\beta} \h{k}_{\alpha\beta\mu;\nu} =
\sum_{k=1}^{n}\T{k}^{\alpha\beta} \h{k}_{\alpha\beta\nu;\mu} ,
\end{equation}
valid for any family of symmetric tensors $\T{k}^{\alpha\beta}$ such
that $\sum_{k=1}^{n}\T{k}^{\alpha\beta}\h{k}_\beta{}^\mu$ is also
symmetric in $\alpha$ and $\mu$.

It is important to emphasize that these combinatorial formulas for
the perturbations of the curvature tensors at a general order are
extremely useful for computational purposes. Essentially, the
problem of perturbations is reduced to that of listing the sorted
partitions of a given number, which can be done very fast in any
computer-algebra system. This fact simply reflects the recursive
differential origin of the perturbation process. The formulas in
this section have been implemented in the free package {\em xPert},
briefly described in Appendix \ref{computer}.


\section{Spherical background}
\label{sphericalsymmetry}

Following GS \cite{GeSe79, GeSe80}, the manifold of a spherical
spacetime $(M^4, g_{\mu\nu})$ is described as a product of the form
$M^4=M^2\times S^2$ using a coordinate system $x^\mu=(x^A, x^a)$
adapted to the $S^2$ orbits of spherical symmetry. The two scalars
$x^A=\{x^0, x^1\}$ provide coordinates for the 1+1 Lorentzian
manifold $M^2$ with boundary, whereas $x^a=\{x^2\equiv\theta,
x^3\equiv\phi\}$ are the usual spherical coordinates on the sphere
$S^2$. The 4-dimensional metric $g_{\mu\nu}$ and the energy-momentum
tensor $t_{\mu\nu}$ can always be written
\begin{eqnarray} \label{sphericalgdecomposition}
&& g_{\mu\nu}(x^D, x^d)dx^\mu dx^\nu = \\
&& \quad g_{AB}(x^D)dx^A dx^B + r^2(x^D)\, \gamma_{ab}(x^d)dx^a dx^b
, \nonumber \\ \label{sphericaltdecomposition}
&& t_{\mu\nu}(x^D, x^d)dx^\mu dx^\nu = \\
&& \quad t_{AB}(x^D)dx^A dx^B + \frac{1}{2} r^2(x^D)\, Q(x^D)\,
\gamma_{ab}(x^d)dx^a dx^b, \nonumber
\end{eqnarray}
where $r$ is a scalar field on $M^2$, $g_{AB}$ is a Lorentzian
metric tensor on $M^2$, and $\gamma_{ab}$ is the round (unit
Gaussian curvature) metric on $S^2$. Using this decomposition it is
straightforward to express all 4-dimensional curvature tensors in
terms of the curvature tensors of $g$ and $\gamma$ and the
derivatives of the scalar $r$, as shown in Ref. \cite{GeSe79}. In
doing so, one usually introduces the vector field
\begin{equation}
v_A \equiv \frac{r_{,A}}{r} = (\log r)_{,A}
\end{equation}
to avoid working with logarithms of $r$. Covariant derivatives of a
vector $f$ on the manifolds $M^4$, $M^2$ and $S^2$ will be denoted as
$f_{\mu;\nu}$, $f_{A|B}$ and $f_{a:b}$, respectively. For instance,
the Schwarzschild spacetime has been described along these lines in
Refs. \cite{MaPo05,NaRe05,SaTi01}.

The totally antisymmetric tensors are denoted with the symbol
$\epsilon$ and obey the conventions
$\epsilon_{ABcd}=\epsilon_{AB}\epsilon_{cd}$ with $\epsilon_{01}=1$
on $M^2$ and $\epsilon_{23}=1$ on $S^2$. Note that some authors
interchange the uppercase/lowercase index conventions, and others
use the opposite sign for $\epsilon_{01}$.

On the other hand, as remarked by Newman and Penrose \cite{NePe66},
it is convenient to introduce a basis of complex vectors on $S^2$:
\begin{equation} \label{mvectors}
m^a = \frac{1}{\sqrt{2}}(e_\theta{}^a + i e_\phi{}^a),
\quad {\rm and} \quad
\bar{m}^a = \frac{1}{\sqrt{2}}(e_\theta{}^a - i e_\phi{}^a),
\end{equation}
where $e_\theta{}^a$ and $e_\phi{}^a$ are the unit norm (with
respect to the round metric) basis vectors. From these vectors we
obtain
\begin{equation} \label{mbarm}
\bar{m}^a m^b = \frac{1}{2} ( \gamma^{ab} + i \epsilon^{ab}) ,
\end{equation}
a relation which can be inverted to get
\begin{eqnarray}
\gamma^{ab} &=& m^a \bar{m}^b + \bar{m}^a m^b, \\
\epsilon^{ab} &=& i ( m^a \bar{m}^b - \bar{m}^a m^b ) .
\end{eqnarray}
These vectors are null, $\gamma_{ab} m^a m^b = \gamma_{ab} \bar{m}^a
\bar{m}^b = 0$, and are normalized so that $\gamma_{ab} m^a
\bar{m}^b = 1$.


\section{Tensor harmonics}
\label{harmonicssection}

The theory of tensor harmonics on the sphere $S^2$ is a very well
known subject \cite{Edmo60,Galindo,Thorne}. In this section we
recast it into the GS notation, which is particularly useful for the
type of algebraic computations that must be performed in high-order
perturbation theory. After a brief review of the GS harmonics we
show that, in order to handle second and higher orders of
perturbation, it is convenient to work with harmonics that posses an
increasingly high number of indices. A generalization of the GS
harmonics to many indices is then defined and shown to be closely
related to the Wigner rotation matrices (also known as spin-weighted
harmonics in General Relativity \cite{Goldberg}). Finally, we give
closed formulas for arbitrary products of these harmonics. Appendix
\ref{harmonics} explains our choice of conventions.

\subsection{GS notation for harmonics}

An orthonormal basis of functions on the sphere $S^2$ is given by
the spherical harmonics $Y_l^m(\theta,\phi)$, which are defined as
the eigenfunctions of the differential operators
\begin{eqnarray}
\gamma^{ab}\nabla_a\nabla_b \, Y_l^m &=& - l(l+1) Y_l^m, \\
\qquad
i\partial_\phi Y_l^m &=& - m Y_l^m ,
\end{eqnarray}
where $\gamma^{ab}$ is the inverse of the round metric on $S^2$, the
angle $\phi$ is defined by choosing a fixed $z$-axis, and $l$ and
$m$ are integers such that $l\geq |m|$. These harmonics are
normalized so that
\begin{equation}
\int d\Omega \; Y_{l'}^{m'} {Y_l^m}^* = \delta_{l'l}\,
\delta_{m'm},
\end{equation}
with $d\Omega$ being the area element on $S^2$ ($d\Omega=\sin\theta
\,d\theta \,d\phi$).

From them, the Regge-Wheeler (RW) \cite{ReWh57} basis of vector
fields on the sphere can be defined as follows. The basis is formed
by the vector fields $Y_l^m{}_{:a}$ and their orthogonal fields
$S_l^m{}_a \equiv \epsilon_{ab}\gamma^{bc}Y_l^m{}_{:c}$.
A basis for 2-tensor fields on the sphere can also be constructed in
a similar way \cite{Zeri70b} and it is formed by three types of
objects: pure-trace tensors can be decomposed using $\gamma_{ab}
Y_l^m$; antisymmetric tensors can be expanded using $\epsilon_{ab}
Y_l^m$; and finally symmetric traceless tensors can be expanded
using
\begin{eqnarray}
Z_l^m{}_{ab}& \equiv& (Y_l^m{}_{:ab})^{\rm TF} = Y_l^m{}_{:ab}
+\frac{l(l+1)}{2}\gamma_{ab} Y_l^m, \\
X_l^m{}_{ab} &\equiv& S_l^m{}_{(a:b)}.
\end{eqnarray}
Indices in round brackets are symmetrized, and the superscript ${\rm
TF}$ denotes the trace-free part. Note that GS use
$2S_l^m{}_{(a:b)}$ instead of $X_l^m{}_{ab}$, different by a factor
of 2.

\subsection{Products of harmonics}
\label{leibnitz}

In the next section we will expand the metric perturbations
$\pert{n}{h}_{\mu\nu}$ in tensor harmonics. From expressions like
(\ref{generalgn}), (\ref{Gamma4}), and (\ref{pertRiemann2}) it is
clear that we need to compute products of several tensor harmonics
when working beyond linear perturbation theory. Even though those
expressions contain products of many harmonics, the problem can be
dealt with recursively because the product of two tensor harmonics
can be decomposed as a series of tensor harmonics of adequate rank.
In principle, we might conclude that at perturbation order $n$ we
need to work with tensor harmonics of rank $2n$ or similar, but the
situation turns out to be simpler in General Relativity.

The formalism starts from perturbations of the metric, which contain
tensor harmonics on $S^2$ of rank zero, one or two, and computes the
decomposition in harmonics of the perturbations of the Einstein
tensor, which also contain harmonics of those ranks. On the other
hand, only second (at most) derivatives of the metric perturbations
will appear in the perturbations of any curvature tensor, at any
order. Finally, as long as we are interested just in perturbations
of curvature tensors, only those contractions in Eq.
(\ref{pertRiemann2}) are required. From these three observations we
conclude that we only need harmonics with up to four indices (if one
works in RW gauge, to be defined below, only three-index harmonics
are required) and formulas for their thirteen products
\begin{equation}
\begin{array}{llll}
YY', & & & \\
YY'{}_{:a}, &\quad Y_{:a}Y'{}_{:b}, & & \\
YY'{}_{:ab}, &\quad Y_{:a}Y'{}_{:bc}, &\quad Y_{:ab}Y'{}_{:cd}, \\
YY'{}_{:abc}, &\quad Y_{:a}Y'{}_{:bcd} &\quad Y_{:ab}Y'{}_{:bcd},
&\quad Y_{:abc} Y'{}_{:def} , \\
YY'{}_{:abcd}, &\quad Y_{:a}Y'{}_{:bcde}, &\quad Y_{:ab}
Y'{}_{:cdef}, &
\end{array}
\end{equation}
where the prime denotes that $Y$ and $Y'$ have different labels $l$
and $m$. Only seven of those are really independent because using
the Leibnitz rule we have relations like
\begin{equation}
Y_{:ab} Y'{}_{:cd} = (Y_{:a}Y'{}_{:cd})_{:b} - Y_{:a} Y'{}_{:cdb}.
\end{equation}
Therefore, computing the expansion formula for the canonical
products $YY'{}_{:a_1...a_n}$ with $n=0,...,6$ would be enough to
solve a general problem of nonspherical perturbations in General
Relativity.

That method would be, however, rather complicated to program, because
it requires to expand the products of multiple harmonics in a very
particular order, and difficult to use in any mathematical proof
involving products of harmonics.
It is far more interesting and general to follow a different
route: we first generalize the GS harmonics to an arbitrary number
of indices and then find a general formula for the product of any
two of them. This has two important advantages: first, it is more
efficient and simple for our algebraic code because all cases are
considered in a single formula. Second, the formalism is more general:
it can be
applied to arbitrary matter models, it is possible to perturb
objects like derivatives of the Riemann tensor, or can be used in
other problems (for example theories of gravity with more than two
derivatives in their basic equations).

\subsection{Higher-order generalization of GS tensors}

Complete bases for second and higher-order tensors can be easily
constructed.
There always exist two nontrivial symmetric trace-free (STF) tensors
\begin{eqnarray} \label{Zhigh}
\!Z_l^m{}_{a_1...a_s}\!& \equiv& \!(Y_l^m{}_{:a_1...a_s})^{\rm STF}
=-\epsilon_{(a_1}{}^b X_l^m{}_{ba_2...a_s)}, \\ \label{Shigh}
\!X_l^m{}_{a_1...a_s} \!&\equiv& (S_l^m{}_{a_1:a_2...a_s})^{\rm STF}
=\epsilon_{(a_1}{}^b Z_l^m{}_{ba_2...a_s)},
\end{eqnarray}
valid for $|m|\le l$ and $1\le s\le l$. In all other cases the
harmonics are defined to be identically zero, except for $s=0$, when
$Z_l^m \equiv Y_l^m$. Note that, in fact, we do not need
symmetrization on the far right-hand side because the tensors $Z$
and $X$ are traceless. All other objects in the basis can be
obtained from products of $\gamma$, $\epsilon$ and the basis for
tensors of order $s-2$. For example the basis for 3-index tensors is
given by $Z_l^m{}_{abc}$, $X_l^m{}_{abc}$, and six independent
combinations of $\gamma_{ab}Z_l^m{}_{c}$, $\gamma_{ab}X_l^m{}_{c}$,
$\epsilon_{ab}Z_l^m{}_{c}$, $\epsilon_{ab}X_l^m{}_{c}$, and their
index-permutations. The general case results from the iteration of
the relations (valid for $s\ge 2$):
\begin{eqnarray}
\label{dZ}
Z_l^m{}_{a_1...a_s:b} &=& Z_l^m{}_{a_1...a_sb}
+ \frac{(l+s)(l-s+1)}{2} \\
& \times& \left[\frac{1}{2}\gamma_{(a_1a_2}Z_l^m{}_{a_3...a_s)b}
-\gamma_{b(a_1}Z_l^m{}_{a_2...a_s)}\right] \nonumber , \\
\label{dX}
X_l^m{}_{a_1...a_s:b} &=& X_l^m{}_{a_1...a_sb}
+ \frac{(l+s)(l-s+1)}{2} \\
& \times& \left[\frac{1}{2}\gamma_{(a_1a_2}X_l^m{}_{a_3...a_s)b}
-\gamma_{b(a_1}X_l^m{}_{a_2...a_s)}\right] \nonumber .
\end{eqnarray}
Appendix \ref{STF} gives a different approach to expand the
definitions (\ref{Zhigh}) and (\ref{Shigh}).

Remembering the definitions of the scalars $Z_l^m\equiv Y_l^m$ and
$X_l^m\equiv 0$, and those of the vectors $Z_l^m{}_a\equiv
Y_l^m{}_{:a}$ and $X_l^m{}_a\equiv S_l^m{}_a$, we obtain the three
remaining special cases:
\begin{eqnarray}
\label{dZ0}
Z_l^m{}_{:a} &=& Z_l^m{}_a , \\
\label{dZ1}
Z_l^m{}_{a:b} &=& Z_l^m{}_{ab} - \frac{l(l+1)}{2}\gamma_{ab}
Z_l^m , \\
\label{dX1}
X_l^m{}_{a:b} &=& X_l^m{}_{ab} - \frac{l(l+1)}{2}\epsilon_{ab}
Z_l^m .
\label{Xlast}
\end{eqnarray}
Formulas (\ref{dZ})--(\ref{dX1}) for the STF tensors $Z$ and $X$
constitute a complete set of simplification rules which allow us to
express any derivative of a tensor harmonic field on the sphere
in a unique canonical way.
Note that $\epsilon_{ab}$ appears only in Eq. (\ref{dX1}).

Finally, it is important to point out that all harmonics have a
well-defined parity under inversion of axes. This is because the
parity of $\gamma_{ab}$, $\epsilon_{ab}$, and the scalar harmonics
$Y_l^m$ is +1, -1 and $(-1)^l$, respectively, and because taking
covariant derivatives does not change the parity. It is usual to
separate harmonics with momentum $l$ into two families: tensors
$Z_l^m{}_{a_1...a_s}$ have parity $(-1)^l$ (these are said to be
``polar'' or of ``even'' polarity), and tensors
$X_l^m{}_{a_1...a_s}$ have parity $(-1)^{l+1}$ (``axial'' or ``odd''
polarity). One must not confuse parity and polarity: whereas all
equations must have a well-defined parity at any order in
perturbation theory, polarity is only useful in the first-order
theory because products of harmonics couple the two polarities, as
we will see.

\subsection{A formula for the product of tensor harmonics}

The product of two scalar harmonics can be expanded in terms of
finite sums of scalar harmonics using Clebsch-Gordan coefficients
\cite{Edmo60}:
\begin{equation}\label{scalars}
Y_{l'}^{m'}\, Y_l^m = \sum_{l''=|l'-l|}^{l'+l}
\E{0}{l'}{m'}{0}{l}{m}{l''} \, Y_{l''}^{m+m'},
\end{equation}
where we have defined the symbol
\begin{equation}\label{scalE}
\E{0}{l'}{m'}{0}{l}{m}{l''} \equiv
 \sqrt{\frac{(2l+1)(2l'+1)}{4\pi(2l''+1)}}
\, \C{l'}{m'}{l}{m}{l''}{m'+m}\C{l'}{0}{l}{0}{l''}{0}.
\end{equation}
Recall that the Clebsch-Gordan coefficient $\C{l'}{0}{l}{0}{l''}{0}$
vanishes if $l'+l+l''$ is odd. This fact guarantees that
only scalars with parity $(-1)^{l''}=(-1)^{l'+l}$ are present in the
expansion.

In this subsection we will construct a generalization of Eq.
(\ref{scalars}) valid for any pair of tensor harmonics on the
2-sphere. There are two main routes to find such a formula. The
standard route, explored in Appendix \ref{pureorbitalharmonics} and
followed by most books in Quantum Mechanics, is adapted to the
3-dimensional Euclidean structure of $R^3$ and uses the so-called
``pure-orbital'' harmonics ${\cal O}^{j,m}_l{}_{i_1...i_s}$, which
transform under a representation of ``total angular momentum'' $j$,
with $|m|\le j$, and whose Cartesian components are eigenfunctions
of the ``orbital angular momentum'' operator (\ref{L2scalar}) with
eigenvalue $l(l+1)$. This latter property becomes very useful when
solving wave equations in a 3-dimensional setting. Unfortunately,
these harmonics are not transverse to the radial direction, a fact
that unnecessarily complicates the analysis of the radiation in the
far region.

Here we will follow the second route, based on the so-called
``pure-spin'' harmonics ${\cal Y}^{\pm s, m}_l{}_{a_1...a_s}$. They
are adapted to the 2-sphere, and hence are transverse to the radial
direction. Besides, they are closely related to the Wigner
representation matrices of the rotation group, for which a product
formula is well known. These harmonics can be defined in the
following way.

Let us consider the unit sphere,
with points described in a certain frame by coordinates
$(\theta',\phi')$. A rotation $R$ of that frame assigns new
coordinates $(\theta,\phi)$ to the same physical points, so that
scalar fields $f$ transform as
\begin{equation}
f'(\theta,\phi) = f(\theta',\phi') \equiv D(R) f(\theta,\phi) .
\end{equation}
Since spherical harmonics provide an irreducible representation of
the rotation group, their transformation rule must have the form
\begin{eqnarray}
D(R)Y_l^m(\theta,\phi) &=& Y_l^m(\theta',\phi') \\
&=& \sum_{m'=-l}^{l} {\cal D}_{m'm}^{(l)}(\alpha,\beta,\gamma)
Y_l^{m'}(\theta,\phi), \nonumber
\end{eqnarray}
where $\alpha$, $\beta$, and $\gamma$ are the three Euler angles
corresponding to the rotation $R$. Explicit formulas for the
components of the matrices ${\cal
D}_{m'm}^{(l)}(\alpha,\beta,\gamma)$ have been computed by Wigner
\cite{Edmo60} and are reproduced in Appendix \ref{harmonics}. Most
important for us, the product of any two of these matrices for the
same rotation $R$ can be expanded using Clebsch-Gordan coefficients:
\begin{eqnarray} \label{calDproduct}
&& {\cal D}_{m_1'm_1}^{(j_1)}(R)\; {\cal D}_{m_2'm_2}^{(j_2)}(R) =
\\ &&
\sum_{j} \C{j_1}{m_1}{j_2}{m_2}{j}{m_1+m_2}
\C{j_1}{m_1'}{j_2}{m_2'}{j}{m_1'+m_2'}
{\cal D}_{m_1'+m_2',m_1+m_2}^{(j)}(R). \nonumber
\end{eqnarray}

Making use of these matrices, the spin-weighted scalar harmonics can
be defined as \cite{Goldberg}
\begin{equation}\label{Spin-wei}
{}_sY_{l,m}(\theta,\phi) \equiv \sqrt{\frac{2l+1}{4\pi}}
{\cal D}_{-s,m}^{(l)}(0,\theta,\phi) .
\end{equation}
In particular ${}_0Y_{l,m}=Y_l^m$.

STF tensors on the unit sphere can be constructed from the vectors
$m^a$ and $\bar{m}^a$ defined in Eq. (\ref{mvectors}). Two
independent STF tensors of rank $s$ are
\begin{equation}
m^{a_1}...m^{a_s} \qquad {\rm and} \qquad
\bar{m}^{a_1}...\bar{m}^{a_s},
\end{equation}
because $\gamma_{ab}m^am^b=0=\gamma_{ab}\bar{m}^a\bar{m}^b$. We then
define the pure-spin tensor harmonics with $s\ge 0$ indices on $S^2$
as
\begin{eqnarray}\label{Yharmo}
{\cal Y}_l^{s,m}{}_{a_1...a_s}\! &\equiv & \!\!(-1)^s k(l,s)
{\cal D}^{(l)}_{s,m}(0,\theta,\phi)  m_{a_1}...m_{a_s} , \\
\!\!\!{\cal Y}_l^{-s,m}{}_{a_1...a_s} \!&\equiv & k(l,s)  {\cal
D}^{(l)}_{-s,m}(0,\theta,\phi) \bar{m}_{a_1}...\bar{m}_{a_s} ,
\label{Yharmo2}
\end{eqnarray}
with
\begin{equation}\label{kcoeffic}
k(l,s) = \sqrt{\frac{(2l+1)(l+s)!}{ \,\,2^{s+2}\,\pi\,(l-s)!}}.
\end{equation}
The normalization factors are introduced so that the GS harmonics
are (for $s\ge 1$)
\begin{eqnarray}\label{Zharmo}
Z_l^m{}_{a_1...a_s} &=& {\cal Y}_l^{s,m}{}_{a_1...a_s} +{\cal
Y}_l^{-s,m}{}_{a_1...a_s}, \quad \\
\label{Zharmo2}-iX_l^m{}_{a_1...a_s} &=& {\cal
Y}_l^{s,m}{}_{a_1...a_s} -{\cal Y}_l^{-s,m}{}_{a_1...a_s}.
\end{eqnarray}
For the case $s=0$, one has $Z_l^m={\cal Y}^{0,m}_l=Y_l^m$.

We can invert the previous relations to get (except for the special
case $s=0$)
\begin{eqnarray}
{\cal Y}^{s,m}_l{}_{a_1...a_s} &=& \left(m_{a_1}\bar{m}^b\,
Y_l^m{}_{:ba_2...a_s}\right)^{\rm STF} , \\
{\cal Y}^{-s,m}_l{}_{a_1...a_s} &=& \left(\bar{m}_{a_1}m^b\,
Y_l^m{}_{:ba_2...a_s}\right)^{\rm STF}.
\end{eqnarray}
Formula (\ref{calDproduct}) provides the following product of
pure-spin harmonics with the same sign:
\begin{eqnarray} \label{productY}
&& {\cal Y}_{l'}^{\pm s',m'}{}_{a_1...a_{s'}}
{\cal Y}_l^{\pm s,m}{}_{b_1...b_{s}} = \\
&& \sum_{l''=|l-l'|}^{l'+l} \E{\pm s}{l}{m}{\pm s'}{l'}{m'}{l''}
{\cal Y}_{l''}^{\pm (s'+s),m'+m}{}_{a_1...a_{s'}b_1...b_s} ,
\nonumber
\end{eqnarray}
where we have introduced the real coefficients
\begin{equation}
\E{s'}{l'}{m'}{s}{l}{m}{l''} \equiv
\frac{k(l',|s'|)k(l,|s|)}{k(l'',|s+s'|)} \C{l'}{m'}{l}{m}{l''}{m'+m}
\C{l'}{s'}{l}{s}{l''}{s'+s},
\end{equation}
which generalize the coefficients (\ref{scalE}). These inherit from
the Clebsch-Gordan coefficients the symmetry properties
\begin{eqnarray}
& \E{-s}{l}{m}{-s'}{l'}{m'}{l''}=\E{s}{\,\,l}{-m}{s'}{\,\,l'}{-m'}
{l''}=
(-1)^{l'+l-l''}\, \E{s}{l}{m}{s'}{l'}{m'}{l''} , & \quad
\label{Eproperties1} \\
& \E{s}{l}{m}{s'}{l'}{m'}{l''} = \E{s'}{l'}{m'}{s}{l}{m}{l''} .
\label{Eproperties2}
\end{eqnarray}
From the fact that $\C{l}{m}{l}{m}{l''}{2m}=0$ for odd $l''$ we also
get that the $E$-coefficients vanish for odd $l''$ if $l=l'$ and
either $m=m'$ or $s=s'$.

For the remaining products of pure-spin harmonics (those with
opposite signs), we obtain (assuming e.g. that $s'\ge s$ without
loss of generality)
\begin{eqnarray}
&& {\cal Y}_{l'}^{\mp s',m'}{}_{a_1...a_{s'}}
{\cal Y}_l^{\pm s,m}{}_{b_1...b_{s}} = \\
\!\!\!&& \!\sum_{l''=|l-l'|}^{l'+l}\!\! \E{\pm s}{l}{m}{\mp
s'}{l'}{m'}{l''}  {\cal Y}_{l''}^{\mp (s'-
s),m'+m}{}_{a_{s+1}...a_{s'}} T^{\pm s}{}_{a_1b_1...a_sb_s},
\nonumber
\end{eqnarray}
where the products
\begin{eqnarray}
T^{s}{}_{a_1b_1...a_sb_s}&\equiv&(-1)^s
\bar{m}_{a_1}m_{b_1}...\bar{m}_{a_s}m_{b_s} , \\
T^{-s}{}_{a_1b_1...a_sb_s}&\equiv&(-1)^s
m_{a_1}\bar{m}_{b_1}...m_{a_s}\bar{m}_{b_s}
\end{eqnarray}
must be expanded using Eq. (\ref{mbarm}). We define $T^0\equiv 1$.

Adopting the notation ${}^{\mbox{\tiny $(\pm)\!$}}{\cal Z}\equiv
{\cal Y}^s \pm {\cal Y}^{-s}$ for all $s$ (where we have obviated
the rest of sub and superindices), and introducing the tensors
${\cal T}^\pm\equiv \frac{1}{2}(T^{-s}\pm T^{s})$ and the
alternating sign $\varepsilon\equiv(-1)^{l+l'-l''}$, we hence arrive
at the final formula (assuming again that $s'\ge s$) valid for the
product of any two generalized harmonics
\begin{eqnarray}\label{Zproduct}
&& {}^{\mbox{\tiny $(\sigma')\!$}}{\cal
Z}_{l'}^{m'}{}_{a_1...a_{s'}}
{}^{\mbox{\tiny $(\sigma)\!$}}{\cal Z}_l^m{}_{b_1...b_s} = \\
&& \sum_{l''=|l'-l|}^{l'+l} \E{s}{l}{m}{s'}{l'}{m'}{l''}
\;{}^{\mbox{\tiny $(\varepsilon\sigma\sigma')\!$}}
{\cal Z}_{l''}^{m'+m}{}_{a_1...a_{s'}b_1...b_s} \nonumber \\
&&+\!\!\sum_{l''=|l'-l|}^{l'+l} \!\!\sigma
\E{-s}{l}{m}{s'}{l'}{m'}{l''} \left( {}^{\mbox{\tiny
$(\varepsilon\sigma\sigma')\!$}} {\cal
Z}_{l''}^{m'+m}{}_{a_{s+1}...a_{s'}}
{\cal T}^{+}_{a_1b_1...a_sb_s}\right. \nonumber \\
&&  +\left.{}^{\mbox{\tiny $(-\varepsilon\sigma\sigma')\!\!$}} {\cal
Z}_{l''}^{m'+m}{}_{a_{s+1}...a_{s'}} {\cal T}^{-}_{a_1b_1...a_sb_s}
\right), \nonumber
\end{eqnarray}
which constitutes the main result of this section. The first sum in
this formula is very simple [similar to that in Eq.
(\ref{productY})] and involves only harmonics with $s'+s$ indices.
The second sum involves harmonics with $s'-s$ indices and has a more
complicated structure in order to include the case of products with
scalar harmonics.


\section{Nonspherical perturbations}

\subsection{2+2 decomposition}

From now on, we adopt the following 2+2 decomposition of the
perturbations of the metric $g_{\mu\nu}$ and the energy-momentum
tensor $t_{\mu\nu}$:
\begin{widetext}
\begin{eqnarray}\label{metricdecomposition}
\Delta^n[g_{\mu\nu}] \equiv
\h{n}_{\mu\nu} \equiv \sum_{l,m}
\left(
\begin{array}{cc}
\pert{n}{H}{}_l^m\!{}_{AB} \; Z_l^m &
\pert{n}{H}{}_l^m\!{}_A \; Z_l^m{}_b
+\pert{n}{h}{}_l^m\!{}_A \; X_l^m{}_b \\
\pert{n}{H}{}_l^m\!{}_A \; Z_l^m{}_b +\pert{n}{h} {}_l^m\!{}_A \;
X_l^m{}_b & \;\pert{n}{K}_l^m \; r^2\gamma_{ab} Z_l^m +
\pert{n}{G}_l^m \; r^2Z_l^m\!{}_{ab} + \pert{n}{h}_l^m \;
X_l^m\!{}_{ab}
\end{array}
\right),
\end{eqnarray}
\begin{eqnarray}\label{emdecomposition}
\Delta^n[t_{\mu\nu}] \equiv
\pert{n}{\psi}_{\mu\nu} \equiv \sum_{l,m}
\left(
\begin{array}{cc}
\pert{n}{\Psi}{}_l^m\!{}_{AB} \; Z_l^m &
\pert{n}{\Psi}{}_l^m\!{}_A \; Z_l^m{}_b
+\pert{n}{\psi}{}_l^m\!{}_A \; X_l^m{}_b \\
\pert{n}{\Psi}{}_l^m\!{}_A \; Z_l^m{}_b +\pert{n}{\psi}{}_l^m\!{}_A
\; X_l^m{}_b & \; \pert{n}{\tilde\Psi}_l^m \; r^2\gamma_{ab} Z_l^m +
\pert{n}{\Psi}_l^m \; Z_l^m\!{}_{ab} + \pert{n}{\psi}_l^m \;
X_l^m\!{}_{ab}
\end{array}
\right).
\end{eqnarray}
\end{widetext}
The polar (axial) components of the perturbations are denoted with
uppercase (lowercase) letters. All indices have been displayed in
this expression, forcing us to use up to five indices in the tensor
perturbation $\pert{n}{H}{}_l^m\!{}_{AB}$. For $n=1$ these
expansions in harmonics (and in particular the choice of factors
$r^2$) reduce to those of Ref. \cite{GeSe80}, with some changes in
the notation and the mentioned difference of normalization of the
axial tensor $X_l^m{}_{ab}$.

\subsection{Gauge dependence}

The perturbations (\ref{metricdecomposition}) and
(\ref{emdecomposition}) are not invariant under changes of gauge,
i.e., under changes of the point-to-point identification between the
perturbed and unperturbed spacetimes. Those changes can be thought
of as diffeomorphisms on one of those manifolds. Sonego and Bruni
\cite{SoBr98} have given a complete description of the Taylor
expansion of families of diffeomorphisms around the identity map. A
given family is described by an infinite collection of vector fields
$\pert{n}{\xi}^\mu$ --one at each order in the power series
expansion-- in terms of which the action of the diffeomorphisms can
be expressed using Lie derivatives. For example, up to second order
the result is
\begin{eqnarray}
\overline{\Delta[T]}- \Delta[T] &=& {\cal L}_{\pert{1}{\xi}} T,
\\
\overline{\Delta^2[T]}- \Delta^2[T] &=& {\cal L}_{\pert{2}{\xi}} T \\
&+& 2 {\cal L}_{\pert{1}{\xi}} \Delta[T]+ {\cal L}^2_{\pert{1}{\xi}}
T,\nonumber
\end{eqnarray}
for any tensor field $T$, where the overline denotes a different
choice of gauge. In particular, for the metric field,
\begin{eqnarray}
\overline{\h{1}_{\mu\nu}}-\h{1}_{\mu\nu} &=&
{\cal L}_{\pert{1}{\xi}}g_{\mu\nu} , \\
\overline{\h{2}_{\mu\nu}}-\h{2}_{\mu\nu} &=& {\cal
L}_{\pert{2}{\xi}} g_{\mu\nu}\\
&+& 2 {\cal L}_{\pert{1}{\xi}} \h{1}_{\mu\nu} + {\cal
L}^2_{\pert{1}{\xi}}g_{\mu\nu}. \nonumber
\end{eqnarray}
These formulas allow us to compute the perturbations in any desired
gauge from their values in a particular one.

In principle, it may be possible to form gauge-invariant
combinations of the perturbations at any desired order. For
simplicity, nonetheless, here we will perform our calculations by
imposing a particular choice of gauge: the generalization to higher
orders ($n>1$) of the RW gauge, namely,
\begin{equation}
\pert{n}{H}{}_l^m\!{}_A = 0 , \quad
\pert{n}{G}{}_l^m = 0 , \quad
\pert{n}{h}{}_l^m = 0 .
\end{equation}
This leads to a full metric $\tilde{g}_{\mu\nu}$ whose components
$Ab$ and $ab$ obey four local gauge conditions at all points,
\begin{equation} \label{generalRW}
\tilde{g}_{Ab:c}\,  g^{bc}=0, \qquad
\tilde{g}_{ab} = \tilde{K} g_{ab},
\end{equation}
for some generic scalar field $\tilde{K}$ on $M^4$. To be more
precise, the RW conditions select the point-to-point mapping between
the perturbed and unperturbed spacetimes in such a way that the
pull-back $\tilde{g}$ of the perturbed metric into the background
manifold obeys Eq. (\ref{generalRW}). Conversely, using these four
conditions it is easy to see that it is always possible to impose
the RW gauge at all perturbation orders in any family of metrics, at
least at a local level, so that the gauge is well posed.

\subsection{Evolution equations}

The spherical background metric (\ref{metricdecomposition})
satisfies the Einstein equations $G_{\mu\nu} = 8\pi\, t_{\mu\nu}$,
which can be decomposed as \cite{GeSe79}
\begin{eqnarray}
G_{AB} &=& -2(v_{A|B}+v_Av_B) \\
&+& g_{AB}\left(-\frac{1}{r^2}+2v_C{}^{|C}+3v_Cv^C\right)
= 8\pi \, t_{AB} , \nonumber \\
G_a{}^a &=& -\sign{2}{R} + 2v_A{}^{|A}+ 2v_A v^A = 8\pi \, Q .
\end{eqnarray}

The evolution of the first-order perturbations is sche\-matically
given by the six GS equations
\begin{eqnarray}\label{GS1}
E_{AB}[\pert{1}{h}_l^m] &=& 8\pi \pert{1}{\Psi}_l^m{}_{AB},
\\\label{GS2}
E_A[\pert{1}{h}_l^m] &=& 8\pi \pert{1}{\Psi}_l^m{}_A, \\\label{GS3}
\tilde{E}[\pert{1}{h}_l^m] &=& 8\pi \pert{1}{\tilde\Psi}_l^m, \\
\label{GS4}
E[\pert{1}{h}_l^m] &=& 8\pi \pert{1}{\Psi}_l^m, \\\label{GS5}
O_A[\pert{1}{h}_l^m] &=& 8\pi \left(
\pert{1}{\psi}_l^m{}_A -\frac{1}{2} Q \h{1}_l^m{}_A\right), \\
\label{GS6}
O[\pert{1}{h}_l^m] &=& 8\pi \pert{1}{\psi}_l^m,
\end{eqnarray}
where $\pert{1}{h}_l^m$ represents the 10 first-order perturbations
with labels $l$ and $m$, and the $E$ and $O$ differential operators
are given in Appendix \ref{operators}. The $E$ operators contain
only polar metric perturbations and the $O$ operators only axial
ones.

The evolution of the second-order perturbations is dictated by the
same equations, except for that now the left-hand side contains
extra sources that are quadratic in the first-order perturbations:
\begin{eqnarray}
\label{eq2A} \!\!E_{AB}[\pert{2}{h}_l^m] \!&+&
\!\!\sum_{\bar{l},\hat{l}} \sum_{\bar{m},\hat{m}}
\Source{\varepsilon}{\bar{l}}{\bar{m}}{\hat{l}}{\hat{m}}{l}{m}{}_{AB}
\!=\! 8\pi \pert{2}{\Psi}_l^m{}_{AB}, \\
\label{eq2B} E_A[\pert{2}{h}_l^m] &+& \sum_{\bar{l},\hat{l}}
\sum_{\bar{m},\hat{m}}
\Source{\varepsilon}{\bar{l}}{\bar{m}}{\hat{l}}{\hat{m}}{l}{m}{}_A
= 8\pi \pert{2}{\Psi}_l^m{}_A, \\
\label{eq2C} \tilde{E}[\pert{2}{h}_l^m] &+& \sum_{\bar{l},\hat{l}}
\sum_{\bar{m},\hat{m}}
\tildeSource{\varepsilon}{\bar{l}}{\bar{m}}{\hat{l}}{\hat{m}}{l}{m}
= 8\pi \pert{2}{\tilde\Psi}_l^m, \\
\label{eq2D} E[\pert{2}{h}_l^m] &+& \sum_{\bar{l},\hat{l}}
\sum_{\bar{m},\hat{m}}
\Source{\varepsilon}{\bar{l}}{\bar{m}}{\hat{l}}{\hat{m}}{l}{m}
= 8\pi \pert{2}{\Psi}_l^m, \\
\label{eq2E} O_A[\pert{2}{h}_l^m] &-& i \sum_{\bar{l},\hat{l}}
\sum_{\bar{m},\hat{m}}
\Source{-\varepsilon}{\bar{l}}{\bar{m}}{\hat{l}}{\hat{m}}{l}{m}{}_A = \\
&& 8\pi \left( \pert{2}{\psi}_l^m{}_A - \frac{1}{2} Q \h{2}_l^m{}_A
\right), \nonumber \\
\label{eq2F} O[\pert{2}{h}_l^m] &-& i \sum_{\bar{l},\hat{l}}
\sum_{\bar{m},\hat{m}}
\Source{-\varepsilon}{\bar{l}}{\bar{m}}{\hat{l}}{\hat{m}}{l}{m}{} =
8\pi \pert{2}{\psi}_l^m,
\end{eqnarray}
with the usual restrictions on the values of $\bar{m}$ and
$\hat{m}$, and with both $\bar{l}$ and $\hat{l}$ being independent
and running over all nonnegative integers. The structure of the
sources is rather peculiar, owing to the mixture of polarities that
appears in the product of harmonics. This fact is encoded in the
{\it polarity sign} $\sigma$ of the sources
$\Source{\sigma}{\bar{l}}{\bar{m}}{\hat{l}}{\hat{m}}{l}{m}$, which
is always given in terms of the associated sign $\varepsilon\equiv
(-1)^{\bar{l}+\hat{l}-l}$ and thus completely determined for each
term of the sum, so that it cannot be chosen freely. Sources with
polarity sign $\sigma=+1$ contain terms $polar\times polar$ and
$axial\times axial$ with real coefficients. Sources with polarity
sign $\sigma=-1$ contain terms of the form $polar \times axial$ with
purely imaginary coefficients. This form ensures an adequate
behavior of the equations under complex conjugation (that we denote
with the symbol $*$). In particular, using (\ref{Eproperties1}), we
have for all sources and for all $\bar{l}, \hat{l}, l$:
\begin{eqnarray}
\left[\Source{\varepsilon}{\bar{l}}{\bar{m}}{\hat{l}}
{\hat{m}}{l}{m}\right]^{*} &=&
\Source{\varepsilon}{\bar{l}}{-\bar{m}}{\hat{l}}
{-\hat{m}}{l}{-m}{},
\\ \left[\Source{-\varepsilon}
{\bar{l}}{\bar{m}}{\hat{l}}{\hat{m}}{l}{m}\right]^{*}&=& -
\;\Source{-\varepsilon}{\bar{l}}{-\bar{m}}{\hat{l}}
{-\hat{m}}{l}{-m}{}.
\end{eqnarray}
The sign $\varepsilon$ alternates when any of the $l$ labels
changes. Therefore, all equations have generically both types of
sources.

On the other hand, we see that some pairs of equations share the
sources: for example, Eqs. (\ref{eq2D}) and (\ref{eq2F}) alternate
their sources $\sign{+}{S}$ and $\sign{-}{S}$ for particular sets of
labels $\lred, \lblue, l$. The same thing happens with the pair
(\ref{eq2B}) and (\ref{eq2E}). The operators $E_{AB}$ and
$\tilde{E}$, however, have their own pair of sources. As a result,
we need to compute eight sources in total, instead of twelve.

Using the expansion (\ref{pertRiemann2}) and the definition of the
metric perturbations (\ref{metricdecomposition}), we can expand the
Einstein equations at second order assuming, without loss of
generality, that there are only two first-order perturbations, but
allowing these to be completely arbitrary, in particular assigning
arbitrary harmonic labels to them.
From now on the coefficients and harmonic labels of those two
perturbations will be denoted as $\hat{h}$ and $\bar{h}$, with all
other perturbation amplitudes vanishing.
In this way, we avoid to deal
with sums that include (quadratic) couplings between an infinite
number of first-order perturbations. The expansion contains many
terms with products of tensor harmonics: we count 1275, 972 and 1347
source terms in $\Delta[G_{AB}]$, $\Delta[G_{Ab}]$ and
$\Delta[G_{ab}]$, respectively (still at the 2+2 abstract level,
without any expansion in coordinate ranges). These products of
harmonics must then be expanded using formula (\ref{Zproduct}). We
now analyze the sources separately.

The source of $\Delta[G_{AB}]$ contains products of harmonics of the
form $\bar{Z}^{abc}\hat{Z}_{abc}$, $\bar{Z}^{ab}\hat{X}_{ab},$ etc.
(harmonics with four indices do not appear in RW gauge). The final
expression can be rearranged to arrive at the sources:
\begin{widetext}
\begin{eqnarray}
\label{sourcerealAB}
\Source{+}{\lred}{\mred}{\lblue}{\mblue}{l}{m}{}_{AB} &=&
-\frac{2}{r^4} \E{2}{\lred}{\mred}{-2}{\lblue}{\mblue}{l}\, g_{AB}
\hred_C \hblue^C
+\frac{\lred(\lred+1)}{r^2}\frac{\lblue(\lblue+1)}{r^2}
\E{0}{\lred}{\mred}{0}{\lblue}{\mblue}{l}
\left(\hred_A\hblue_B-g_{AB}\hred_C\hblue^C\right) \\
&+& \frac{1}{r^2} \E{1}{\lred}{\mred}{-1}{\lblue}{\mblue}{l}
\bigg\{4\hred^C( \hblue_{(A|B)C} -\hblue_{C|AB} +\hblue_C v_{A|B}
+2\hblue_{C|(A}v_{B)}) +4\hred^C{}_{|C}\hblue_{(A|B)}
-2\hred^C{}_{|A}\hblue_{C|B}
\nonumber \\
&-&2\hblue_A{}^{|C}\hred_{B|C} + g_{AB} \bigg[
2\hred^{(C|D)}\hblue_{(C|D)} -\hred^C\hblue^D
\bigg(12\frac{r_{|CD}}{r}-4v_Cv_D\bigg)
-2r^{-4}(r^2\hred^C)_{|C}(r^2\hblue^D)_{|D} \nonumber \\
&+&\bigg(2\;\sign{2}{R}-4\frac{\lred^2+\lred-1}{r^2}
+\frac{4}{3}\frac{(r^3)^{|D}{}_{D}}{r^3}\bigg) \hred^C\hblue_C
+\frac{4}{r^2}\epsilon^{CD}\hred_C(r^4\Pblue)_{|D} -2r^4\Pred\Pblue
+ 4r^2\epsilon^{CD}\hblue_C v_D\Pred \bigg]\bigg\} \nonumber\\
&+& \frac{1}{2r^2} \E{1}{\lred}{\mred}{-1}{\lblue}{\mblue}{l}
\left[2\Hred^C{}_C\Hblue_{AB}-4\Hblue_{AC}\Hred^C{}_B
+g_{AB}\left(4\Kred\Kblue+3\Hred^{CD}\Hblue_{CD}-\Hred^C{}_C
\Hblue^D{}_D\right)
\right] \nonumber \\
&+& \E{0}{\lred}{\mred}{0}{\lblue}{\mblue}{l} \bigg\{
-\frac{\lred^2+\lred+\lblue^2+\lblue-2}{r^2}\Hred_{AB}\Kblue
+\Hblue^C{}_{AB}\Kred_{|C} +\frac{2}{r^3} \Hred_{AB}
(r^3\Kblue_{|C})^{|C} + (\Kred\Kblue)_{|AB} - \Kred_{|A}\Kblue_{|B}
\nonumber \\
&+& 4 v_{(A} \Kred_{|B)} \Kblue
+\frac{1}{2}\Hred_{CD(A}\Hblue^C{}_{B)}{}^D
-\Hred^{CD}\left[r^{-2}(r^2\Hblue_{CAB})_{|D}-\Hblue_{CD|BA}\right]
-\frac{1}{2}\Hred_{DAB}\Hblue^{DC}{}_C
\nonumber \\
&+& (\Hred_{AB}-g_{AB}\Hred^F{}_F)\bigg[ \Hblue^C{}_C{}^{|D}{}_D -
\Hblue^{CD}{}_{|CD} +2
\Hblue^C{}_{C|D}v^D-4\Hblue^{CD}{}_{|C}v_D-2\Hblue^{CD}
(2v_{C|D}+3v_Cv_D)\nonumber  \\
&+&\Hblue^C{}_C\bigg(\frac{\sign{2}{R}}{2}-\frac{\lblue^2+\lblue}
{r^2}\bigg)\bigg]+ g_{AB} \bigg[
\frac{\lred^2+\lred}{r^2}(\Hred^C{}_C+2\Kred)\Kblue
-\Hblue^{DC}{}_C\Kred_{|D}-\frac{2}{r^3}\Hred^{CD}
(r^3\Kblue_{|C})_{|D}-\frac{2}{r^2}\Kred\Kblue
\nonumber \\
&-&\frac{1}{r^3}[r^3(\Kred\Kblue)_{|C}]^{|C}
+\frac{3}{2}\Kred_{|C}\Kblue^{|C}
+2\Hblue^{FDE}[(\Hred_{FC}v^C-\Hred^C{}_Cv_F)g_{DE}+ v_F\Hred_{DE}]
+\frac{1}{4}\Hred_{FC}{}^{C}\Hblue^{FD}{}_D
\nonumber  \\
&-&\frac{1}{4}\Hred_{CDF}\Hblue^{CDF}+(\Hred^{CD}\Hblue_C{}^F
-\Hred^C{}_C\Hblue^{DF})\bigg[g_{DF}\bigg(\frac{\lblue^2
+\lblue}{r^2}-\frac{\sign{2}{R}}{2}\bigg)+2(2v_{D|F}+3v_Dv_F)
\bigg]\bigg]\bigg\}, \nonumber\\
\label{sourceimagAB}
\Source{-}{\lred}{\mred}{\lblue}{\mblue}{l}{m}{}_{AB} &=&
\frac{2i}{r^2} \E{1}{\lred}{\mred}{-1}{\lblue}{\mblue}{l}
\bigg\{\Hred_{AB}\hblue^C{}_{|C} + \Hred^C{}_C \hblue_{(A|B)} - 2
\Hred^C{}_{(A}\hblue_{B)|C} + 2 (\Hred_{AB|C} - \Hred_{C(A|B)})
\hblue^C \\ &+& g_{AB}
\bigg[\Hred^{CD}\frac{1}{r^2}(r^2\hblue_C)_{|D} - \Hred^C{}_C
\hblue^D{}_{|D} + 2 (\Hred^C{}_{D|C} - \Hred^C{}_{C|D} - \Kred_{|D})
\hblue^D \bigg] \bigg\}.\nonumber
\end{eqnarray}
We have employed that the sums in $\lblue$ and $\lred$ are symmetric
to simplify the form of the sources. Although each individual source
$\Source{\sigma}{\lblue}{\mblue}{\lred}{\mred}{l}{m}$ is not
symmetric under the interchange $(\lblue,\mblue) \leftrightarrow
(\lred,\mred)$, their sum is symmetrized. We have also tried to
simplify the expressions as much as possible by using the GS scalar
$\Pi=\epsilon^{AB}(r^{-2}h_A)_{|B}$. In addition, we have defined
$H_{ABC} \equiv H_{AB|C}+H_{AC|B}-H_{BC|A}$.

On the other hand, the source of $\Delta[G_{Ab}]$ can be decomposed
as
\begin{eqnarray}
\Source{+}{\lred}{\mred}{\lblue}{\mblue}{l}{m}{}_{A} &=&
\frac{2}{r^2} \E{-1}{\lred}{\mred}{2}{\lblue}{\mblue}{l}
\left\{(\hred_{[A|B]}+\hred_Bv_A)\hblue^B -\hred^B\hblue_{(A|B)}
\right\} \\
&+& \E{0}{\lred}{\mred}{1}{\lblue}{\mblue}{l} \bigg\{
\frac{1}{2}\Hred_{BC|A}\Hblue^{BC} +
\Hred^{BC}(\Hblue_{BC|A}-\Hblue_{AB|C}- \Hblue_{BC}v_A) +
\frac{1}{2}(\Hred^B{}_{B|C}-2\Hred^B{}_{C|B})\Hblue_A{}^C
\nonumber \\
&+& (\Kred\Kblue)_{|A} + \frac{1}{2}\Kred_{|A}\Hblue^B{}_B
+\frac{\lred^2+\lred}{r^2}\bigg[3(\hred_{[A|B]}+\hred_Bv_A)
\hblue^B-\hred^B\hblue_{(A|B)}
+r^2\hred_A(r^{-2}\hblue^B)_{|B}\bigg] \bigg\} ,
\nonumber\\
\Source{-}{\lred}{\mred}{\lblue}{\mblue}{l}{m}{}_{A} &=&
\frac{-i}{r^2} \E{2}{\lred}{\mred}{-1}{\lblue}{\mblue}{l} \left\{
\Hred_{AB}\hblue^B+\hred^B\Hblue_{AB}\right\}
\\
&+& \frac{i}{2} \E{1}{\lred}{\mred}{0}{\lblue}{\mblue}{l} \bigg\{
\frac{\lblue^2+\lblue}{r^2}\left[-\Hred_{AB}\hblue^B+\hred^B
\Hblue_{AB}+(\Hred^B{}_B-2\Kred)\hblue_A-\hred_A(\Hblue^B{}_B
+2\Kblue)\right]+ 2\hred^B{}_{|A}\Kblue_{|B}
\nonumber \\
&-& 2 r^{-2} (r^2\hred^B)_{|B} \Kblue_{|A} - 2 r^{-2} \hred^B (r^2
\Kblue_{|B})_{|A} +\frac{2}{r^2} \hred_A \bigg(
2\tilde{E}\left[\h{1}_\lblue^\mblue\right] +
\Kblue(r^2\;\sign{2}{R}-\lred^2-\lred)+ (r^2\Kblue)^{|B}{}_B\bigg)
\nonumber \\
&+& 2r^2\epsilon^{BC}\Pred\Hblue_{AB|C}
-r^2\epsilon_{AB}\Pred(\Hblue_C{}^{C|B}-2\Hblue^{BC}{}_{|C})
+2r^{-2}\epsilon_{AB}(r^4\Pred)_{|C}\Hblue^{BC} \bigg\} .
\nonumber\end{eqnarray}

Finally, the manipulations for $\Delta[G_{ab}]$ are more
complicated, involving up to 3091 terms in some intermediate steps.
The resulting expression can be organized in the following four
sources:
\begin{eqnarray}
\tildeSource{+}{\lred}{\mred}{\lblue}{\mblue}{l}{m} &=&
\frac{1}{r^2}\E{-1}{\lred}{\mred}{1}{\lblue}{\mblue}{l} \bigg\{
\Hred^{AB}\Hblue_{AB}-\frac{1}{2}\Hred^A{}_A\Hblue^B{}_B
+2\hred^{A|B}\hblue_{A|B}-\frac{2}{r^2}(r\hred^A)_{|A}(r\hblue^B)_{|B}
\\
&+&2\hred^A\hblue^B\bigg[2R_{AB}+3v_Av_B-g_{AB}
\frac{\lblue^2+\lblue-1-2rr^{|C}{}_{|C}}{r^2} \bigg]
+\frac{4}{r}\epsilon^{AB}\hred_A(r^3\Pblue)_{|B} \bigg\}\nonumber
\\
&+& \frac{1}{2}\E{0}{\lred}{\mred}{0}{\lblue}{\mblue}{l} \bigg\{
-\frac{\lred^2+\lred}{r^2}\frac{\lblue^2+\lblue}{r^2}
\hred^A\hblue_A +
2(\Hred^A{}_A-\Kred)(\Hblue^{BC}{}_{|BC}-\Hblue^B{}_B{}^{|C}{}_C)
\nonumber \\
&+&\frac{1}{2}\Hred^{AB|C}(2\Hblue_{AC|B}-3\Hblue_{AB|C})
+\frac{1}{2}(2\Hred^A{}_{B|A}-\Hred^A{}_{A|B})
(2\Hblue^{CB}{}_{|C}-\Hblue^C{}_C{}^{|B})\nonumber \\
&+&\Hred^{AB}\bigg[\Hblue_{AB}
\left(\frac{\lblue^2+\lblue}{r^2}-{}^{(2)}R\right) -2 v_B
\Hblue^C{}_{C|A}+4(\Hblue_{AC}v^C)_{|B}
+\frac{4}{r}(r\Hblue_{AC})^{|C}v_B -2\Hblue_{AB|C}v^C\bigg]
\nonumber \\
&-&2 [\Hred^{AB}r^{-2}(r^2\Kblue)_{|B}]_{|A} + \Hred^A{}_{A|B}
r^{-2}(r^2\Kblue)^{|B} +\sign{2}{R}\, \Hblue^A{}_A \Kred - 4
\Hred^{AB}\Kblue v_A v_B - \Kred^{|A}\Kblue_{|A} \bigg\},
\nonumber\\
\tildeSource{-}{\lred}{\mred}{\lblue}{\mblue}{l}{m} &=&
\frac{2i}{r^2} \E{-1}{\lred}{\mred}{1}{\lblue}{\mblue}{l} \bigg\{
\left(g^{AB}\Hred^C{}_C-\Hred^{AB}\right)\hblue_{A|B}+
\left(\Kred_{|B}+4\Hred^A{}_{[A|B]}-\Hred^A{}_Av_B\right)
\hblue^B\bigg\},
\\
\Source{+}{\lred}{\mred}{\lblue}{\mblue}{l}{m} &=&
\E{1}{\lred}{\mred}{1}{\lblue}{\mblue}{l} \bigg\{ \frac{1}{2}
\Hred^{AB}\Hblue_{AB} + \Hred^A{}_A \Kblue + r^4 \Pred \Pblue -
2\frac{\lred^2+\lred-1}{r^2}\hred^A \hblue_A \bigg\}
+\E{0}{\lred}{\mred}{2}{\lblue}{\mblue}{l} \Hred^{AB}\Hblue_{AB} ,
\\
\Source{-}{\lred}{\mred}{\lblue}{\mblue}{l}{m} &=& -2i
\E{1}{\lred}{\mred}{1}{\lblue}{\mblue}{l} (\Kred\hblue^A)_{|A} -i
\E{0}{\lred}{\mred}{2}{\lblue}{\mblue}{l} \left\{
2(\Hred^{AB}\hblue_A)_{|B}-\Hred^A{}_{A|B}\hblue^B\right\}.
\end{eqnarray}
\end{widetext}

In spite of the obvious increase of complexity from the first to the
second-order equations, we want to stress that the final expressions
given above are still manageable and fully general, except for the
choice of RW gauge. They can be particularized to the case of any
spherical background, dynamical or not, containing any type of
matter, and expressed in any kind of background coordinates
(polar-radial, null, comoving, etc.)

In situations with just a single first-order perturbation we will
have $\hat l=\bar l$ and $\hat m=\bar m$. In these circumstances,
the $E$-coefficients vanish for odd $l$. This implies that $\hat
l+\bar l-l$ is always even and therefore the sources $\sign{-}{S}$
are never excited in the polar equations (\ref{eq2A})--(\ref{eq2D}),
neither are the sources $\sign{+}{S}$ in the axial equations
(\ref{eq2E}) and (\ref{eq2F}). In particular, $\sign{-}{S}_{AB}$ and
$\sign{-}{\tilde{S}}$ are never excited. This has been the case
encountered in many previous investigations in second-order
perturbation theory. For instance, for the perturbations of a slowly
rotating star a single axial $l=1$ mode has been assumed \cite{CF91},
and the studies of the close-limit black hole collisions have
considered a single polar $l=2$ first-order perturbative mode
\cite{GNP96*}. It will be very interesting to
discuss what kind of interactions between modes are excited through
the other types of sources, an issue that we plan to analyze in
future investigations.

\subsection{Energy-momentum conservation}

A complete set of evolution equations is obtained only after
specifying the particular type of matter content of the system
(including as such the vacuum). Some simple systems like scalar
fields of perfect fluids are completely defined dynamically by
energy-momentum conservation, but this is not the case in general.
However, we can generally analyze the consequences of perturbing the
matter conservation equations, as well as use this analysis as a
check of the perturbed Bianchi identities, and hence as a
consistency check of the sources given in the previous subsection.

Like every object in perturbation theory, the energy-momentum
conservation law can be expanded into a hierarchy of linear
equations, all sharing the principal part,
\begin{equation}\label{emconservation}
\Delta^n[T_{\mu\nu}{}^{;\nu}]=0.
\end{equation}
At zeroth order we use the decomposition
(\ref{sphericaltdecomposition}) provided by the background, which
leads to a nontrivial relation
\begin{equation}
Q \, v_A=\frac{1}{r^2}\left(r^2 t_{AB}\right)^{|B}.
\end{equation}
Given its vectorial character, the energy-momentum conservation
equation can be decomposed into three geometric parts at higher
orders: a vector equation in the polar sector and two scalar (one
polar and one axial) equations. At first order, those equations can
be written in compact notation as
\begin{eqnarray}
L_A[\pert{1}{\psi}_l^m,\pert{1}{h}_l^m]=0, \label{1orderenergy1} \\
L[\pert{1}{\psi}_l^m,\pert{1}{h}_l^m]=0, \\
\tilde L[\pert{1}{\psi}_l^m,\pert{1}{h}_l^m]=0.
\label{1orderenergy2}
\end{eqnarray}
The operators $L_A$, $L$, and $\tilde{L}$ are defined in Appendix
\ref{operators}.

The second-order perturbation adopts the same form with additional
quadratic sources:
\begin{eqnarray}\label{order2equations1}
L_A[\pert{2}{\psi}_l^m,\pert{2}{h}_l^m] &+& \sum_{\bar{l},\hat{l}}
\sum_{\bar{m},\hat{m}}
\source{\varepsilon}{\bar{l}}{\bar{m}}{\hat{l}}{\hat{m}}{l}
{m}{}_{A}=0, \\
L[\pert{2}{\psi}_l^m,\pert{2}{h}_l^m] &+& \sum_{\bar{l},\hat{l}}
\sum_{\bar{m},\hat{m}}
\source{\varepsilon}{\bar{l}}{\bar{m}}{\hat{l}}
{\hat{m}}{l}{m}{}_{}=0, \\\label{order2equations2} \tilde
L[\pert{2}{\psi}_l^m,\pert{2}{h}_l^m] &+& \sum_{\bar{l},\hat{l}}
\sum_{\bar{m},\hat{m}}
\source{-\varepsilon}{\bar{l}}{\bar{m}}{\hat{l}}{\hat{m}}{l}{m}{}_{}=0.
\end{eqnarray}
Such sources can be computed starting from Eq.
(\ref{emconservation}), decomposing it using formulas
(\ref{metricdecomposition}) and (\ref{emdecomposition}), and finally
applying the tools that we have developed to deal with products of
harmonics. The result, with the same notation used for the sources
of the main equations, is
\begin{widetext}
\begin{eqnarray}
\source{+}{\bar{l}}{\bar{m}}{\hat{l}}{\hat{m}}{l}{m}{}_{A} &=&
\frac{1}{r^2}\E{1}{\lred}{\mred}{-1}{\lblue}{\mblue}{l} \bigg\{
2r^2\psired^B(r^{-2}\hblue_B)_{|A} -\Hblue^B{}_B\Psired_A
-\frac{r^4}{2}Q(r^{-4}\hred_B\hblue^B)_{|A}
+2(\psired_A\hblue^B)_{|B}\\\nonumber&+&
r^2t_A{}^B(r^{-2}\hred^C\hblue_C)_{|B}-2(\hred^C\hblue^Bt_{AB})_{|C}
-r^2t_{BC} (r^{-2}\hred^B\hblue^C)_{|A} \bigg\}\\
\nonumber&+& \frac{1}{2} \E{0}{\lred}{\mred}{0}{\lblue}{\mblue}{l}
\bigg\{ \frac{2\lred(\lred+1)}{r^2}\Kblue\Psired_A +2\Psiblue_A{}^B
\Kred_{|B} +2Qr\Kblue(r^{-1}\Kred)_{|A} -2\Kred\Kblue^{|B}t_{AB}
-2r^2\Psitildeblue(r^{-2}\Kred)_{|A}\\
\nonumber&-&2\Hred^{BC}t_{AB}\Kblue_{|C} -\Psired^{BC}\Hblue_{BC|A}
+\Psired_A{}^B\Hblue^C{}_{C|B} -\frac{2}{r^2}
(r^2\Psired_{AB}\Hblue^{BC})_{|C}
\\&+&\Hred^{BC}\bigg[4\Hblue_C{}^D{}_{|(A}
t_{D)B}-\Hblue^D{}_{D|C}t_{AB}-\Hblue_{BC|D}t_A{}^D
+\frac{2}{r^2}(r^2\Hblue_{BD}t_A{}^D)_{|C}\bigg]
\bigg\},\nonumber\\
\source{-}{\bar{l}}{\bar{m}}{\hat{l}}{\hat{m}}{l}{m}{}_{A} &=&
-\frac{i}{r^2}\E{1}{\lred}{\mred}{-1}{\lblue}{\mblue}{l} \bigg\{
\Hblue^B{}_B\psired_A + \big(2\Hblue^{BC}\hred_C
-\Hblue^C{}_C\hred^B\big)t_{AB} - 2 \hred^B \Psiblue_{AB}
\\ \nonumber
&-& 2 r^2 \Psiblue^B (r^{-2}\hred_B)_{|A} - 2
(\Psiblue_A\hred^B)_{|B} \bigg\},\\
\source{+}{\bar{l}}{\bar{m}}{\hat{l}}{\hat{m}}{l}{m}{}_{}&=&
\frac{1}{r^2}\E{2}{\lred}{\mred}{-1}{\lblue}{\mblue}{l} \bigg\{
2\hblue_A \psired^A +2 \psiblue_A\hred^A +2 (\psired\hblue^A)_{|A}
-\Psired \Hblue^A{}_A - Q\hred^A\hblue_A - 2 \hred^A\hblue^B t_{AB}
\bigg\}\nonumber
\\&+&\frac{1}{r^2}\E{0}{\lred}{\mred}{1}{\lblue}{\mblue}{l}
\bigg\{ \lred(\lred+1)\left[\psiblue_A\hred^A-\psired_A\hblue^A -
\hblue^A\hred^B t_{AB} + \frac{Q}{2}\hblue^A\hred_A\right] +
(\lblue-1)(\lblue+2)\Kred\Psiblue \nonumber\\\nonumber
&-&2(r^2\Psiblue_A\Hred^{AB})_{|B} + 2 r^2 \bigg[\Psiblue^A
\Kred_{|A}-\Psitildered\Kblue- \Psitildeblue\Kred + Q\Kred \Kblue +
\frac{1}{2}\Psiblue^B\Hred^A{}_{A|B}\bigg]
\\
&+&r^2\Hblue^{AB}\bigg[2\Hred_{BC}t^C{}_A - \Psired_{AB}
-\frac{Q}{2}\Hred_{AB} + g_{AB}\bigg(\Psitildered -
\frac{Q}{2}\Kred\bigg)\bigg]\bigg\},\nonumber\\
\source{-}{\bar{l}}{\bar{m}}{\hat{l}}{\hat{m}}{l}{m}{}_{}&=&
\frac{i}{r^2}\E{2}{\lred}{\mred}{-1}{\lblue}{\mblue}{l} \left\{ 2
\hred_A \Psiblue^A - 2 \hblue_A \Psired^A - \psired \Hblue^A{}_{A} -
2 (\Psired \hblue^A)_{|A} \right\}\\ \nonumber&+&
\frac{i}{r^2}\E{0}{\lred}{\mred}{1}{\lblue}{\mblue}{l} \bigg\{\lred
(\lred + 1) (\hblue_A \Psired^A - \hred_A \Psiblue^A) + (\lblue -
1)(\lblue + 2) \psiblue \Kred - 2 (r^2 \psiblue_A \Hred^{AB})_{|B} -
2 (r^2\Psitildered\hblue^A)_{|A}\\&+&2 r^2 \psiblue^A \Kred_{|A} +
\Kred (Qr^2\hblue^A)_{|A} + (Qr^2\Hred^{AB}\hblue_B)_{|A} +
r^2\Hred^A{}_{A|B} \left(\psiblue^B - \frac{Q}{2}\hblue^B\right)
\bigg\}.\nonumber
\end{eqnarray}
\end{widetext}


\section{Conclusions}

Perturbation theory has been highly successful in General Relativity
and nowadays still plays a relevant role in the simulations of the
dynamics of many physical and astrophysical systems. On the one
hand, it allows us to interpret certain dynamical processes as the
evolution of a perturbative mode of a simpler process, or as the
interaction among several of such modes. On the other hand, using
perturbation theory we can evolve systems in which very different
physical phenomena are happening simultaneously with so distinct
space/time scales or amplitudes that a numerical simulation would
fail to follow them all with a satisfactory precision. Even so,
one must be aware of the problem of linearization stability when
perturbing a given spacetime in General Relativity, namely, it may
happen that a perturbatively constructed spacetime with parameter
$\epsilon$ does not correspond to an exact family of solutions
$g_{\mu\nu}(\epsilon)$ of the Einstein equations. This problem
becomes more important when perturbing highly symmetric backgrounds
\cite{Moncrief75}.

However, going beyond first-order perturbation theory has not been
feasible until very recently except in very specific situations. The
present work proposes a systematic approach to high-order
perturbation theory in General Relativity, based on the combination
of a good choice of the theoretical formalism employed for the
description of the problem (implementing symmetry reductions,
covariant notation, and other nice features) and the intensive use
of abstract computer algebra to manipulate the enormous expressions
that unavoidably appear in this field.

We have first given a number of formulas which permit to compute
very efficiently and at any order the perturbation of all relevant
curvature tensors in General Relativity, and have implemented them
in the {\em Mathematica} package {\em xPert}. These formulas could
be used in very different areas of gravitational physics, including
theories that depart from standard General Relativity (like in the
case of models with extra dimensions, curvature corrections, or in
braneworld scenarios).

We have then generalized to higher orders the well known GS
formalism for nonspherical first-order perturbations of a spherical
spacetime. This formalism is considered to be optimal for the
perturbative study of a number of astrophysical scenarios of
interest, except for the possible construction of master
scalars and equations referred to in the Introduction. The
generalization of the GS formalism put forward here will make the
perturbative analysis even more powerful, leading to more precise
results and allowing to describe interactions between different
modes.

With this purpose, we have constructed a generalization of the GS
harmonics, which turn out to be very closely related to the Wigner
rotation matrices (spin-weighted harmonics in the General Relativity
community). We have also obtained a general formula for the product
of any pair of them, and implemented all these results in another
package called {\em Harmonics}.

In addition, we have computed all the equations of the generalized
GS formalism at second order (including those of energy-momentum
conservation) and simplified them to a form manageable enough as to
allow to write them down in this paper. These equations are
completely general except for the restriction to a spherical
background: they can be used with any background, dynamical or not,
they can be coupled to any matter model, and they have been given in
covariant form, so that any coordinate system can be used on the
background manifold. We have presented the results in RW gauge,
which is well posed at all perturbative orders. The original GS
formalism also employs gauge-invariant variables, and we are
currently working in this direction, with the aim at determining the
gauge-invariants at second order in full generality.

The second-order equations are essentially the same as the
first-order equations, but they also include complicated quadratic
sources. We have disentangled the structure of these sources and
shown that, in previous investigations considering just a single
first-order perturbative mode, many of such sources were not
excited. We plan to study the role of these new sources in future
work.

Let us conclude remarking that, once the gauge-invari\-ants be
determined, the formalism and our computer-algebra tools
implementing it will be ready to be applied to problems of
astrophysical or conceptual relevance, such as the emission of
gravitational radiation in the collapse of a rotating star, or the
coupling of perturbative modes of a critical spacetime.

\acknowledgments
D.B. acknowledges financial support from the FPI program of the
Regional Government of Madrid.
J.M.M.-G. acknowledges the financial aid provided by the
I3P framework of CSIC and the European Social Fund.
This work was also supported by the Spanish MEC Projects
No. FIS2004-01912 and No. FIS2005-05736-C03-02.


\appendix

\section{Spherical functions}
\label{harmonics}

Several conventions are employed in the literature for the special
functions used in the theory of representations of the 3-dimensional
rotation group. Here we mostly follow the conventions of Edmonds
\cite{Edmo60}.

The spherical harmonics $Y_l^m(\theta,\phi)$ are
\begin{equation} \label{Ydef}
Y_l^m(\theta,\phi) \equiv \sqrt{\frac{(2l+1)(l-m)!}{4\pi\,(l+m)!}}
\, P_l^m(\cos\theta)e^{im\phi},
\end{equation}
where $P_l^m$ is the associated Legendre function
\begin{equation} \label{Pdef}
P_l^m(x) \equiv \frac{(-1)^m}{2^l
l!}(1-x^2)^{m/2}\frac{d^{\,l+m}}{dx^{l+m}} (x^2-1)^l .
\end{equation}
The {\it Mathematica} functions {\tt SphericalHarmonicY} and {\tt
LegendreP} are indeed those defined above.

For a rotation of the reference frame described by the Euler angles
$(\alpha,\beta,\gamma)$, the convention adopted for the unitary
matrix in a representation ${\cal D}^{(l)}$ of $SU(2)$ is
\begin{equation} \label{Ddef}
{\cal D}^{(l)}_{m'm}(\alpha,\beta,\gamma) = e^{im'\alpha}
d^{(l)}_{m'm}(\beta) e^{im\gamma}.
\end{equation}
The $\beta$-transformation is given by
\begin{eqnarray} \label{ddef}
&& d^{(l)}_{m'm}(\beta) = \\
&& \sum_\sigma \frac{(-1)^{l-m'-\sigma}
\sqrt{(l+m)!(l-m)!(l+m')!(l-m')!}}
{(l-m'-\sigma)!\,(l-m-\sigma)!\,(m'+m+\sigma)!\,\sigma!}
\nonumber \\
&& \times\left(\sin\frac{\beta}{2}\right)^{2l-m'-m-2\sigma}
\left(\cos\frac{\beta}{2}\right)^{m+m'+2\sigma} \nonumber
\end{eqnarray}
where the sum ranges over those integers $\sigma$ for which the
arguments of the factorials are all nonnegative.

\section{Symmetric trace-free tensors}
\label{STF}

Given any tensor $T_{i_1...i_l}$ over a vector space of dimension
$d$ with a metric $g_{ij}$, we construct its STF part as
\begin{eqnarray}
&&[T_{i_1...i_l}]^{STF} \equiv \\
&&\quad\sum_{m=0}^{[l/2]} a^{(m)}_{l,d}
g_{(i_1i_2}...g_{i_{2m-1}i_{2m}}
S_{i_{2m+1}...i_l)}{}^{j_1}{}_{j_1...}{}^{j_m}{}_{j_m}\nonumber
\end{eqnarray}
with $S_{i_1...i_l}\equiv T_{(i_1...i_l)}$ and $[l/2]$ the integer
part of $l/2$. The coefficients of the expansion are determined by
the trace-free condition, and are given by
\begin{equation}
a^{(m)}_{l,d} = \frac{l!}{(-4)^m m! (l-2m)!}
\frac{\Gamma[l+d/2-1-m]}{\Gamma[l+d/2-1]} .
\end{equation}
In our case, $d=2$ for the unit sphere. These formulas allow us to
compute any of the $Z_l^m{}_{a_1...a_s}$ in terms of the derivatives
$Y_l^m{}_{:a_1...a_s}$ or viceversa. Note that derivatives of
$Y_l^m$ with indices sorted differently are not equal, but can be
transformed into a term with the desired order of indices plus terms
with a lower number of derivatives.

\section{Pure-orbital harmonics}
\label{pureorbitalharmonics}

In this Appendix, we briefly discuss the formalism of pure-orbital
tensor harmonics and its relation with the pure-spin harmonics.
Firstly, one constructs pure-orbital vector harmonics by composing
scalar harmonics of angular momentum $l$ with a set of 3-dimensional
vectors $t^m{}_i$ which transform under a representation of spin 1
\cite{Galindo}:
\begin{equation}
{\cal O}^{j,m}_{l}{}_i \equiv
\sum_{m'=-1}^{+1}\C{l}{m-m'}{1}{m'}{j}{m} \, Y_l^{m-m'}\,
t^{m'}{}_i,
\end{equation}
with $j=l-1,l,l+1$ and $|m|\le j$.
The vectors $t^m{}_i$ are defined in terms of a fixed orthonormal
Cartesian basis:
\begin{equation}
t^{\pm 1}{}_i = \frac{\mp e_x{}_i-i e_y{}_i}{\sqrt{2}}, \qquad
t^0{}_i = e_z{}_i.
\end{equation}
(The index $i$ is an abstract index on the manifold $R^3$ with
Euclidean metric, in which $S^2$ is embedded.) These vector
harmonics ${\cal O}^{j,m}_{l}{}_i$ transform under a representation
of total angular momentum $j$ and their Cartesian components are
eigenfunctions, with eigenvalue $l(l+1)$, of the $S^2$-Laplacian
(also called orbital angular momentum \cite{Thorne})
\begin{equation}
L^2 \equiv -r^2\vec{\nabla}^2+\partial_r(r^2\partial_r) =
-\gamma^{ab}\nabla_a\nabla_b . \label{L2scalar}
\end{equation}

Pure-orbital vector harmonics are however not transverse to the
radial direction,
\begin{equation}
\hat{r}_i {\cal O}^{\,j,m}_{l}{}_i = - \C{j}{0}{1}{0}{l}{0} \;
Y^m_j.
\end{equation}
Therefore, one must take certain linear combinations canceling their
radial contribution to get the GS harmonics.

Pure-orbital bases for higher-rank tensors can be constructed
recursively from the above vector basis. The basis for STF tensors
with $s$ indices and well-defined spin $s$ can be built by
composition of the bases with $s'$ and $s-s'$ indices (with any
$0<s'<s$) as follows:
\begin{equation}
t^m{}_{i_1...i_s} = \sum_{m'=-s'}^{s'} \C{s'}{m'}{s-s'}{m-m'}{s}{m}
\; t^{m'}_{(i_1...i_{s'}}t^{m-m'}_{i_{s'+1}...i_s)}.
\end{equation}
From this, we construct orbital harmonics with $s$ indices:
\begin{equation}\label{orbharmonics}
{\cal O}^{j,m}_{l}{}_{i_1...i_s} \equiv
\sum_{m'=-s}^{+s}\C{l}{m-m'}{s}{m'}{j}{m} \, Y_l^{m-m'}\,
t^{m'}{}_{i_1...i_s},
\end{equation}
which are normalized so that
\begin{equation}
\int d\Omega \; \big({\cal O}^{j,m}_l{}_{i_1...i_s}\big)^* {\cal
O}^{j',m'}_{l'}{}_{i_1...i_s} = \delta_{ll'}\delta_{jj'}\delta_{mm'}
.\end{equation} The symbol $*$ denotes complex conjugation.

One can obtain the following multiplication rule by using Eq.
(\ref{scalars}) and the formulas available in the literature for the
composition of three angular momenta \cite{Galindo}:
\begin{eqnarray}
&&Y_{l'}^{m'}\, {\cal O}^{j,m}_l{}_{i_1...i_s} =\\
&& \quad\sum_{l''=|l'-l|}^{l'+l}\sqrt{\frac{(2l+1)(2l'+1)(2j+1)}
{4\pi}}\, \C{l}{0}{l'}{0}{l''}{0}\nonumber
\\
&&\times \sum_{j'=l''-l}^{l''+l} W(s,j,l'',l';l,j')
\C{j}{m}{l'}{m'}{j'}{m+m'} {\cal O}^{j',m+m'}_{l''}{}_{i_1...i_s},
\nonumber
\end{eqnarray}
where $W$ is the Racah coefficient \cite{Edmo60}. From this formula
it is possible to compute the product of any two orbital harmonics
using the Leibnitz rule, but here we prefer to show their connection
with the pure-spin harmonics, which obey a much simpler
multiplication formula.

The radial component of the orbital harmonics is
\begin{eqnarray}\label{radorbten}
&& \frac{\hat{r}^{i_s}{\cal O}^{j,m}_l{}_{i_1...i_{s-1}i_s}}
{\sqrt{(2s+1)(2l+1)}}= \\ && \qquad \sum_{l'} \C{l}{0}{1}{0}{l'}{0}
W(s-1,1,j,l;s,l') {\cal O}^{j,m}_{l'}{}_{i_1...i_{s-1}}, \nonumber
\end{eqnarray}
which is a sum with only two contributions, $l'=l\pm 1$.

Pure-spin harmonic tensors, with indices on $S^2$, can be obtained
as linear combinations of the pure-orbital ones:
\begin{equation}
{\cal Y}^{s,m}_{j}{}_{a_1...a_s} = \sum_{l=j-s}^{j+s} \sqrt{2l+1} \,
A_{j,s}^l \, {\cal O}^{j,m}_l{}_{a_1...a_s}.
\end{equation}
Asking them to be orthogonal to the radial direction, we find from
Eq. (\ref{radorbten}) that the coefficients of this expansion obey
the ``second-order'' recursion relation
\begin{eqnarray}
A^{l+1}_{j,s} &=& f(l,j,s) f(l-1,j,s) A^{l-1}_{j,s} , \\
f(l,j,s) &=&\! \sqrt{\frac{(s-l+j)(l+j+1-s)}{(s+l-j+1)(l+j+2+s)}}.
\end{eqnarray}
Since the odd-$l$ coefficients decouple from the even-$l$ ones,
there are two independent solutions. Fixing a normalization constant
and employing definition (\ref{kcoeffic}), we get
\begin{equation} A^{(\pm)}{}^l_{j,s}=
\C{\;l}{\pm s}{\;s}{\,\mp s}{j}{\,0}\frac{k(l,s)}{\sqrt{2l+1}}
\sqrt{\frac{4 \pi}{2j+1}},
\end{equation}
which define, respectively, the harmonics
${\cal Y}^{\pm s,m}_j{}_{a_1...a_s}$.

Finally, using formulas (\ref{orbharmonics}) and
(\ref{calDproduct}), and the relation (\ref{Spin-wei}) (with $s=0$)
between rotation matrices and spherical harmonics, one can express
the pure-spin harmonics as
\begin{eqnarray} {\cal Y}^{\pm s,m}_j{}_{a_1...a_s}&=& k(l,s)
{\cal D}^{(l)}_{\pm s,m}(0,\theta,\phi) \\ &\times& \sum_{m'=-s}^s
\left[{\cal D}^{(s)}_{\pm s,m'}
(0,\theta,\phi)\right]^{*}t^{m'}{}_{a_1...a_s}.\nonumber
\end{eqnarray}

\section{GS operators} \label{operators}

For the sake of completeness, we include in this Appendix the metric
part of the GS equations at first perturbative order and in RW
gauge. This is also the linear part of the
perturbation equations at any order. As it is usually done at first
order, we obviate here the labels $l$, $m$, and $n$ of the
perturbations on $M^2$.
\begin{widetext}
\begin{eqnarray}
E_{AB}[\pert{n}{h}_l^m] &\equiv& \bigg[\frac{(l-1)(l+2)}{2r^2} + 3
v_C v^C + 2 v^C{}_{|C} \bigg] H_{AB} + v_C (H^C{}_{B|A} +
H^C{}_{A|B} - H_{AB}{}^{|C})
\\
&-& (v_B K_{|A} + v_A K_{|B}+ K_{|AB}) + g_{AB}
\bigg[r^{-3}(r^3K^{|C})_{|C}- \frac{(l-1)(l+2)}{2r^2}K
\nonumber \\
&-&\frac{l(l+1)}{2r^2}H^C{}_C + (H^C{}_{C|D}-2H^C{}_{D|C})v^D
- (3v_Cv_D+2v_{C|D})H^{CD} \bigg], \nonumber\\
E_A[\pert{n}{h}_l^m] &\equiv& \frac{1}{2}(H^B{}_Bv_A - H^B{}_{B|A} +
H_A{}^B{}_{|B} - K_{|A}),
\\
\tilde{E}[\pert{n}{h}_l^m] &\equiv& \frac{1}{2}\bigg\{ (H_{AB}
-Kg_{AB})\sign{4}{R}^{AB} - \frac{l(l+1)}{2r^2} H^A{}_A +
H^A{}_{A|B}{}^B
\\ \nonumber
&-& 2 H^{A}{}_{B|A}v^B + H^{A}{}_{A|B}v^B
-H^{AB}{}_{|AB} + K^{|A}{}_A + 2 K_{|A}v^A \bigg\}, \\
E[\pert{n}{h}_l^m] &\equiv& -\frac{1}{2}H^A{}_A, \\
O_A[\pert{n}{h}_l^m] &\equiv& \frac{(l-1)(l+2)}{2r^2} h_A
-\frac{1}{2r^2}\left[r^4\left(\frac{h_A}{r^2}\right)_{|C}
-r^4\left(\frac{h_C}{r^2}\right)_{|A}\right]^{|C}, \\
O[\pert{n}{h}_l^m] &\equiv& h^A{}_{|A}.
\end{eqnarray}

On the other hand, the linear parts of the equations for the
perturbations of the energy-momentum conservation law, also in RW
gauge, are the following:
\begin{eqnarray}\label{emconservation1}
L_A[\pert{n}{\psi}_l^m,\pert{n}{h}_l^m] & \equiv &
-\frac{l(l+1)}{r^2}\Psi_A  - 2 v_{A}\tilde\Psi + \frac{1}{r^2}
(r^2\Psi_{AB})^{|B}
\\ \nonumber &-&
\frac{1}{2}t^{BC}H_{BC|A} - \frac{r^2}{2}Q(r^{-2}K)_{|A}
-\frac{1}{2} t_{AB} H^C{}_{C}{}^{|B} - t_{AB} K^{|B}
+\frac{1}{r^2} (r^2 t_{AB} H^{BC})_{|C}, \\
L[\pert{n}{\psi}_l^m,\pert{n}{h}_l^m] &\equiv& \tilde\Psi
-\frac{(l-1)(l+2)}{2 r^2}\Psi + \frac{1}{r^2}(r^2\Psi^{A})_{|A} -
(K-\frac{1}{2}H^A{}_A)\frac{Q}{2} -\frac{1}{2} H^{AB}t_{AB},
\label{emconservation2} \\
\tilde{L}[\pert{n}{\psi}_l^m,\pert{n}{h}_l^m] &\equiv&
\frac{1}{r^2}(r^2 \psi^A)_{|A} - \frac{(l-1)(l+2)}{2r^2} \psi -
\frac{1}{2r^2}(Q r^2 h^A)_{|A}. \label{emconservation3}
\end{eqnarray}
\end{widetext}

\section{Computer implementation}
\label{computer}

Three {\em Mathematica} packages have been constructed during the
course of this investigation, respectively called {\em xPert}, {\em
Harmonics} and {\em xPertGS}. This Appendix describes what is
included in them. The three packages are based on the package {\em
xTensor} \cite{xTensor} for abstract tensor computations, written by
one of us, and also freely available.

The package {\em xPert} implements the equations of Sec.
\ref{GRpert}. In particular, there is a command {\tt Perturbation}
which plays the role of $\Delta$ and is able to give $n$th-order
perturbations from those at order $n-1$. The command {\tt
Perturbation[{\em expr}, {\em n}]} is recursively computed via {\tt
Perturbation[Perturbation[{\em expr}, {\em n-1}], {\em 1}]}. There
is then the command {\tt GeneralPerturbation[{\em expr}, {\em n}]},
which implements the expansion formulas for the metric
(\ref{metricexpansion}), the inverse metric (\ref{generalgn}), the
Christoffel symbols (\ref{Gamma4}), the Riemann tensor
(\ref{pertRiemann2}), and the Ricci tensor and scalar. The general
expansions are much faster than the recursive procedures, and for
instance it is possible to produce within seconds the expressions of
fifth-order perturbation theory with a small PC.

The package {\em Harmonics} implements all the structures that have
been defined on $S^2$ in Sec. \ref{sphericalsymmetry}, Sec.
\ref{harmonicssection}, and Appendices \ref{harmonics}, \ref{STF},
and \ref{pureorbitalharmonics} of this paper. The commands {\tt
PureOrbital[{\em j}, {\em l}, {\em m}][{\em a, b, ...}]} and {\tt
PureSpin[{\em j}, {\em $\pm 1$}, {\em m}][{\em a, b, ...}]} give,
respectively, any pure-orbital or pure-spin harmonics, both in
abstract form (in terms of the bases $t$ or $m$) or providing their
components in any coordinate or noncoordinate basis. The generalized
GS harmonics $Z$ and $X$ have also been defined, incorporating all
their symmetries and properties (\ref{Zhigh})-(\ref{Xlast}), as well
as the product formula (\ref{Zproduct}).

The package {\em xPertGS} needs the previous two packages and has
three sections, closely following the mathematical structure of this
article:
\begin{enumerate}
\item Spherical background. The reduced manifold $M^2$ is defined
with its metric {\tt g[-A,-B]} (and corresponding derivative {\tt
CD[-A]}). The scalar field {\tt r[]} is also defined. The manifold
$M^4$ is constructed as a product of $M^2$ and $S^2$; all tensors on
$M^4$ can be block-decomposed in their respective parts on those
submanifolds.
\item GS first order. First-order perturbations are defined and
their equations computed using the gauge invariants (the total
execution time is of the order of two minutes in a PC).
\item GS second order. Second-order perturbations are defined and
their equations computed and manipulated in order to simplify them
as much as possible (the total execution time is of the order of two
hours).
\end{enumerate}

The packages {\em xPert} and {\em Harmonics} can be freely
downloaded, under the GNU General Public License, from {\small{\tt
http://metric.iem.csic.es/Martin-Garcia/xAct/xPert}}.




\begin{thebibliography}{99}
\bibitem{gen} More generally, one might consider metrics
$\tilde{g}_{\mu\nu}(\epsilon)$ defined on manifolds ${\cal
M}(\epsilon)$ that are related via an $\epsilon$-family of
isomorphisms.
\bibitem{JLM} One may also understand these expansions as a
hierarchy of fields. At order $n$ in the hierarchy, two fields are
identified if they differ by terms multiplied by $\epsilon^{n+1}$.
In a more algebraic language, the domain of the fundamental
quantities is the quotient ring
$\mathbb{R}(\epsilon)/\epsilon^{n+1}$ (where
$\mathbb{R}(\epsilon)$ is the ring of polynomials in the real
variable $\epsilon$ with real coefficients).
\bibitem{regu} We also assume that the spacetime dependence of the
metric and matter fields is (sufficiently) smooth.
\end{thebibliography}
\end{document}